# Thermodynamics of Ionic Thermoelectrics for Low-Grade Heat Harvesting

Xin Qian[1,*], Zhihao Ma[1,2], Qiangqiang Huang[1], Haoran Jiang[3], and Ronggui Yang[1,2,*]

[1] Department of Engineering Thermophysics, School of Energy and Power Engineering, Huazhong University of Science and Technology, Wuhan 430074, China

[2] State Key Laboratory of Coal Combustion, Huazhong University of Science and Technology, Wuhan, 430074, China

[3] Key Laboratory of Efficient Utilization of Low and Medium Grade Energy (MOE), Tianjin University, Tianjin 300072, China.

*Corresponding authors: xinqian21@hust.edu.cn; ronggui@hust.edu.cn.

## Abstract

More than half of the waste heat rejected into the environment has temperatures lower than 100 °C, which accounts for nearly 85 PWh/year worldwide. Efficiently harvesting low-grade heat could be a promising step toward carbon neutrality. Recent developments of ionic thermoelectrics (i-TE) with giant thermopower have provoked intensive interest in using ions as energy and charge carriers for efficient thermal energy harvesting. However, current literature primarily focuses on improving thermopower only, while the ion transport and thermodynamics affecting the efficiencies have been largely neglected. This review article clarifies the fundamentals of electrochemistry and thermodynamics for developing highly efficient i-TE devices. Two major types of i-TE devices, thermo-ionic capacitors (TIC) and thermogalvanic cells (TGC), are discussed in detail. The article analyzes the methods of enhancing ionic thermopower in the literature by taking an entropic point of view. We also derived modified thermoelectric factor $Z$ for both TICs and TGCs that fully incorporate the dynamics of ion transport and electrochemical reactions. Recent developments of hybrid devices showing improved efficiencies, power density, and multifunctionality are reviewed. Finally, we comment on the remaining challenges and provide an outlook on future directions.

Due to the thermodynamic irreversibility and efficiency limitations, nearly 70% of the energy consumed is dissipated into the environment in the form of heat[1]. Among the waste heat rejection, more than 60% is the low-grade heat with temperatures below 100 °C, corresponding to ~ 85 PWh/year worldwide [2]. It has been a great desire yet a longstanding challenge to harvest low-grade heat efficiently. Conventional heat recovery methods such as organic Rankine cycles require heavy maintenance and face challenges of cost-effectiveness when converting distributed low-grade heat into electricity [3-4]. Compact solid-state thermoelectric generators (TEGs) could be a solution for on-site recovery of distributed low-grade heat [5-6], but the low thermopower (~ 0.1 mV/K) together with the high cost of thermoelectric materials severely limits their applications [7]. For example, nearly 400 pairs of thermoelectric legs connected in series are required to achieve a voltage of only 2 V under a temperature difference of ~ 50 K, which imposes great challenges in achieving reasonable cost-effectiveness. Recent developments in ionic thermoelectric (i-TE) devices such as thermo-ionic capacitors (TICs) and thermogalvanic cells (TGCs) provide a novel method for converting low-grade heat to electricity. The thermopower generated by these electrochemical devices (1~100 mV/K) [7-13] is significantly larger than traditional semi-conductor thermoelectric devices, which provokes intensive interests in various applications including low-grade human body heat and industrial waste heat recovery [9, 14-19]. The TICs are based on the ionic thermodiffusion, also known as the Soret effect or the ionic Seebeck effect. Thermopower of TICs is induced by ionic migration along the temperature gradient, where charged ions accumulate at the electrode-electrolyte interface (Fig. 1a). A thermal voltage thus builds up once the concentration profiles of cations and anions are mismatched. In TGCs, thermopower originates from the temperature dependence of electrochemical potential, also known as the thermogalvanic effect. When the two electrodes are kept at different temperatures, a thermal voltage can be measured as the electrochemical potential difference between the two electrodes (Fig. 1b).

Early research on the ionic thermopower, dates back to the discovery of the Soret effect, also known as the ionic Seebeck effect in the 19th century [20]. To quantify the capability of converting temperature difference into electro-motive force, the Soret thermopower ($S_{td}$) is defined as:

$$S_{td} = -\frac{\phi(T_H) - \phi(T_C)}{T_H - T_C}, \tag{1}$$

where $\phi$ denotes the electric potential, $T_H$ and $T_C$ are the electrode temperatures. In aqueous solutions, $S_{td}$ is only on the order of 10-100 μV/K [21]. It was not until the recent decade that larger Sore thermopowers $S_{td}$ have been achieved. Bonetti *et al.* observed large thermopower ~ 7 mV/K in nonaqueous electrolytes of tetrabutylammonium nitrate with 1-dodecanol as solvents, due to the strong "structure making" effects of tetrabutylammonium ions which enhances the hydrogen bond network formed by solvent molecules nearby[22]. Zhao *et al.* measured a thermopower of 11 mV/K in polyelectrolyte PEO-NaOH [23], in which the cations and polyelectrolyte anions have largely mismatched concentration gradients. Confinement by nanochannels is another method to obtain high thermopower [24-25]. The enhanced thermopower can be induced by the broken local charge neutrality [26] of electrolytes under nanoscale confinement, and the ion transport becomes more unipolar compared with bulk electrolytes. High thermopower has also been observed when electrolytes are confined in ion exchange membranes [13, 27] (5-19 mV/K), layered graphene oxide [28] (9 mV/K), and nanocellulose [11] (24 mV/K). Recently, the thermopower of ionic liquids and polymer gels has been elevated to 17 mV/K in ionic gelatin [7], 38.2 mV/K, and 52.9 mV/K in PVA [29-30], and even 87 mV/K in polyaniline and polystyrene sulfonate [12]. The ultrahigh thermopower of these i-TE capacitors also enabled highly sensitive temperature measurements and processing of thermal signals. For example, Jiang *et al.* demonstrated the application of thermosenstive ionic hydrogel for early fire warning [31]. The mechanical flexibility together with the high thermopower of gel-based i-TEs have enabled promising applications in self-powered health monitoring [32]. Dan *et al.* demonstrated an i-TE gated organic transistor that can convert and amplify temperature changes to digital signals, which can be used in thermography and thermosenstive e-skins [33].

On the other hand, the effective thermopower of thermogalvanic cells remains on the order of a few mV/K. The thermogalvanic effect of redox couples is described by the temperature coefficient ($\alpha$) [34] of electrode potential:

$$\alpha = \frac{E(T_H) - E(T_C)}{T_H - T_C}, \tag{2}$$

where $E(T_H)$ and $(T_C)$ denote the electrode potentials at temperatures $T_H$ and $T_C$, respectively. It is worth noting that the temperature coefficient $\alpha$ has the opposite sign convention compared with the ionic Seebeck coefficient $S_{td}$. Under the same temperature difference, a TGC based on electrolytes with $\alpha < 0$ has the same voltage sign as a p-type thermoelectric material [7]. For the sake of simpler discussion throughout this review, an "effective thermopower" of TGCs can be defined based on a unified sign convention to the Seebeck coefficient:

$$S_{tg} = -\alpha = -\frac{E(T_H) - E(T_C)}{T_H - T_C}, \tag{3}$$

which will be adopted when discussing TGCs for the rest of the paper. Pristine aqueous electrolytes have moderate $S_{tg}$ with absolute values typically below 2 mV/K, such as 1.4 mV/K [35] of $Fe(CN)_6^{3-}/Fe(CN)_6^{4-}$ and -1.3 mV/K [36] of $Fe^{3+}/Fe^{2+}$. $S_{tg}$ can be enhanced by tailoring solvents, especially for ionic liquids and polyelectrolytes. For example, $S_{tg}$ of $Co^{III}(bpy)_3^{3+}/Co^{II}(bpy)_3^{2+}$ in ionic liquids and organic solvent systems reaches -1.5~-2.2 mV/K [37-39]. Recently, a high $S_{tg}$ of $Fe^{3+}/Fe^{2+}$ in acetone of –3.6 mV/K is reported [40], which is nearly 3 times higher than aqueous electrolytes of $Fe^{3+}/Fe^{2+}$. $S_{tg}$ can also be enhanced by introducing redox-inactive additives. For example, Huang *et al.* pointed out that the inactive supporting electrolyte could re-arrange the solvent molecules around redox ions due to the structure-making or breaking effects [41]. Yu *et al.* reported a boosted thermopower from 1.4 mV/K to 3.7 mV/K by adding guanidinium cations ($Gdm^+$) to $Fe(CN)_6^{3-}/Fe(CN)_6^{4-}$ to induce the thermosensitive crystallization [42]. Despite these experimental advancements in improving thermopower $S_{tg}$, the mechanism of tuning $S_{tg}$ remains unclear.

While current research reported great progress in improving ionic thermopower [14-15, 19], fundamentals of thermodynamics, electrochemistry, and ion transport in i-TE devices are yet comprehensively investigated despite the pivotal importance of affecting power density and efficiencies. In TICs, ion transport determines the charging speed and discharging power. During the thermal charging, ions migrate to the electrode-electrolyte interface driven by the temperature gradient, until such thermodynamic driving force is counter-balanced by the electric field. The thermal charging time is not only determined by the heat capacity of the TIC device but also by the thermal mobility of ions. Decreasing the thermal charging time is crucial for reducing unwanted heat conduction across the device and for improving efficiency. However, only a few works have considered finite thermal charging time when evaluating efficiency [23, 43]. Since the ions in TICs are redox-inert, charges cannot transfer across the electrode-electrolyte interfaces. Heat is converted into electrostatic energy by ionic charges stored near the electrode surfaces of TICs. During the discharging process, ions are released through ionic diffusion and migration, and tuning such ionic diffusion is the key to improving power density. Different from solid-state TEGs, the current and cell voltage of TICs decay with time due to the capacitive nature of energy conversion (Fig. 1c).

The TGCs can output current continuously when the two electrodes are maintained at different temperatures. Unlike TICs, electrons can transfer across the electrolyte-electrode interfaces with the redox species as electron donors or acceptors. During the discharging process, reactant species migrate towards the electrode surfaces, being oxidized or reduced by the Faradaic current, and then the products migrate away from the electrode surfaces. At the steady state, the electric current must be balanced by the ionic currents (Fig. 1d). Once the reactant transport is slow, the discharging current will also be suppressed, resulting in small power density and low efficiency. Due to the similarity in the operation of TGCs and TEGs, the thermoelectric factor $Z$ or figure of merit $ZT$ is usually employed to characterize the efficiency [9, 42, 44-51]. However, different from the solid-state thermoelectric materials whose $ZT$ is determined by material properties, the efficiency of TGC is related to the redox reactions at the electrolyte-electrode interfaces. As a result, mass transport and electrochemical properties at the electrode-electrolyte interface could play a significant role in affecting the performance of TGCs, in addition to the electrolyte properties. Therefore, directly applying $ZT$ based on material properties for evaluating the efficiency of TGCs is questionable [10]. We will discuss this issue and re-derive a figure of merit incorporating ion transport and interfacial properties in this Review.

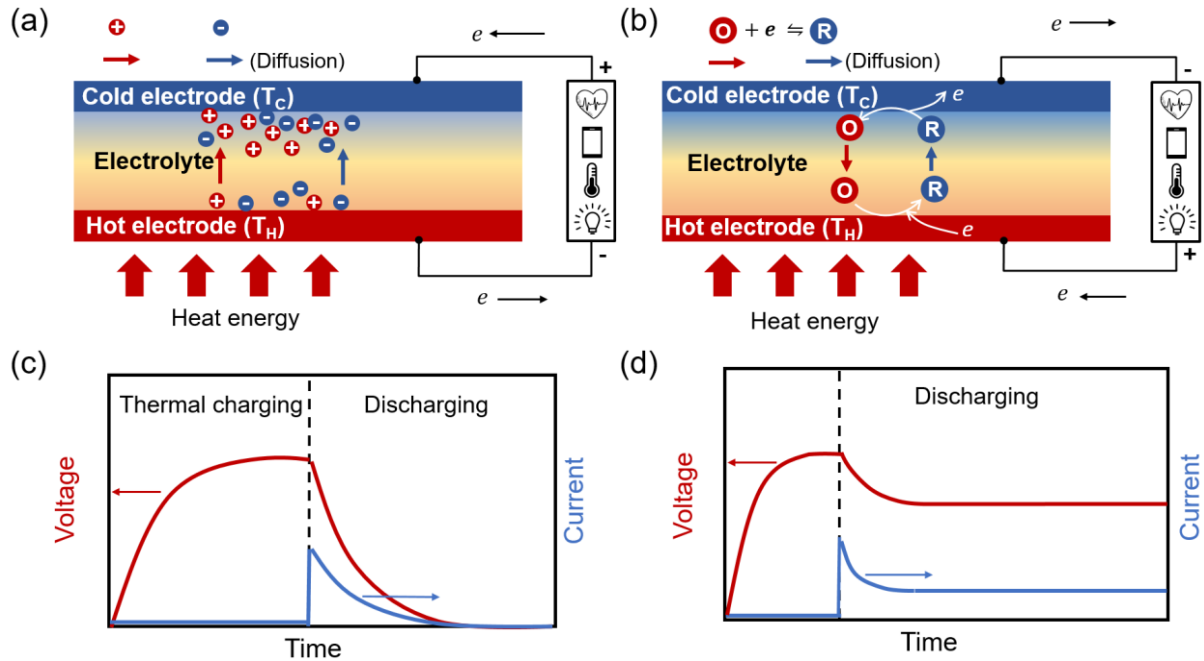

Figure 1. Working principles of (a) TICs and (b) TGCs. Responses of voltage and current during thermal charging and discharge processes of (c) the TICs and (d) TGCs.

This review article aims to provide a comprehensive discussion of the thermopower, power density, and efficiency of i-TE devices from the perspective of thermodynamics. This review begins with discussions on the TIC devices. Fundamental nonequilibrium thermodynamics of ion thermodiffusion is first discussed, and the Soret thermopower $S_{td}$ is shown to be determined by an entropic quantity called Eastman entropy of transfer. Strategies to improve the Soret thermopower $S_{td}$ is then summarized, such as introducing supporting electrolytes and polymer chains with polar functional groups, and the confinement effect of nanochannels. The crucial role of ion transport in affecting the efficiency of TIC devices is discussed, and a modified thermoelectric factor $Z_{TIC}$ incorporating ion transport effects is derived. By analyzing the microscopic ion transport processes using a hopping picture, we show that pursing giant $S_{td}$ might even result in compromised efficiency. To improve the efficiency, TICs based on stacking electrode design are discussed. TGC devices are then comprehensively analyzed. Similar to $S_{td}$, thermogalvanic thermopower $S_{tg}$ can also be understood as the entropy change associated with the energy conversion process, but it is the equilibrium entropy change of the redox reaction that dominantly determines the effective thermopower of TGCs. Methods of improving the thermogalvanic thermopower $S_{tg}$ are reviewed, including solvation shell engineering for improved redox entropy change and introducing activity gradients across the device. The efficiency of TGC devices is shown to be determined by a redefined $Z_{TGC}$ that captures the effects of ion transport and kinetics of electrochemical reactions. We provided strategies for optimizing the electrode and device design of TGCs for high efficiency. Recent developments of the hybrid i-TE devices are summarized, including hybrid devices of TICs and TGCs or solid-state thermoelectric generators for improved thermopower or simultaneous harvesting of temperature gradients and fluctuations, and hybrid devices for electricity-cooling power co-generation and electricity-water co-generation. Finally, we conclude the review with an outlook for future directions.

## ▪ THERMO-IONIC CAPACITORS

The recent discovery of high Soret thermopower provoked intensive interest in harvesting near-room temperature heat, such as using body heat for powering wearable devices [7, 52-53]. However, most electrolytes with giant thermopower (>10 mV/K) have ionic mobilities (~$10^{-8}$ m$^2$/(V·s)) [54-55] much lower than electron/hole mobilities in typical solid-state TE materials such as Bi$_2$Te$_3$ and PbTe (~0.01 m$^2$/(V·s)) [56-57]. As a result, the power density of the current i-TE device is still nearly two orders of magnitude smaller than solid-state TEGS, and further enhancements are still necessary to work with Internet of Things (IoT) devices with higher power requirements such as 3-axis gyroscope (5.3 mW) and magnetic sensor (3.3 mW) [58]. The limited efficiency ($\sim 10^{-5}$-$10^{-2}$ %) [59] is closely related to the slow ion transport. Small thermal mobilities of ions result in a long thermal charging time, during which the unwanted heat conduction is a major source of efficiency loss. In addition to further improving the thermopower (i.e. the electromotive force), overcoming ion transport limitations is also pivotal for boosting the power density and efficiency. This section is organized as follows. We first introduce basic concepts of the thermodiffusion effect and show that the thermopower $S_{td}$ is determined by the mismatch of Eastman entropy of transfer between cations and anions and boundary conditions (open or closed boundaries). We then review recent advancements in improving $S_{td}$, including using nonaqueous solvents, introducing interactions with polymer networks, using ionic liquids or ion complexes, and by confining the electrolytes in nanochannels. The thermodynamics and operation modes of TIC devices will be discussed. Specifically, we derive a refined $Z$-factor analogous to solid-state TEG devices, taking into account the effects of mass transport and thermal coupling with the reservoir. Finally, we review the electrode engineering for optimizing TIC devices and propose a stacking design to improve efficiency.

**Eastman entropy of transfer and Soret thermopower.** The Soret thermopower $S_{td}$ originates from the net charge accumulation at electrode-electrolyte interfaces driven by the temperature gradients. The coupled ionic-thermal transport can be described by Onsager's linear relation:

$$\boldsymbol{J}_i = L_{ii}\left(-\frac{\nabla_T \bar{\mu}_i}{T}\right) + L_{iQ}\nabla\left(\frac{1}{T}\right) \qquad (4)$$

$$J_Q = \sum_i L_{Qi}\left(-\frac{\nabla_T \bar{\mu}_i}{T}\right) + L_{QQ}\nabla\left(\frac{1}{T}\right),$$

where $J_i$ and $J_Q$ are the ionic flux and heat flux, respectively. The thermodynamic driving forces are gradients of electrochemical potential at a constant temperature $\nabla_T \bar{\mu}_i/T$ and gradients of temperature $\nabla(1/T)$. The associated $L_{ii}$, $L_{iQ}$, $L_{Qi}$, $L_{QQ}$ are linear transport coefficients, and $L_{Qi} = L_{iQ}$ according to Onsager reciprocity. By expanding the gradient of electrochemical potential to the concentration gradient ($\nabla n_i$) and electric potential gradient ($\nabla \phi$), i.e. $\nabla_T \bar{\mu}_i = \nabla n_i + z_i e \nabla \phi$, the constitutive equations of coupled ionic-thermal transport can be derived [7, 24]:

$$J_i = -D_i \left(\nabla n_i + n_i \frac{z_i e}{k_B T}\nabla \phi + n_i \frac{Q_i^*}{k_B T^2}\nabla T\right) \quad (5)$$

$$J_Q = \sum_i Q_i^* J_i - k\nabla T, \quad (6)$$

where $D_i = L_{ii} k_B / n_i$ is the diffusion coefficient of ionic species $i$, $k = T^{-2}\left(L_{QQ} - \sum_i \frac{L_{Qi} L_{iQ}}{L_{ii}}\right)$ is the thermal conductivity of the electrolyte. At zero ionic current $J_i$ and electric field $\nabla \varphi$, Eq. (5) can be simplified as:

$$\nabla(\ln n_i) = -\frac{Q_i^*}{k_B T}\nabla(\ln T). \quad (7)$$

It is clear that $Q_i^*$ determines the magnitude of the Soret effect. $Q_i^*$ is known as the heat of transport, defined as:

$$Q_i^* = \frac{L_{Qi}}{L_{ii}}. \quad (8)$$

The quantity $Q_i^*$ can be understood by setting the temperature gradient to zero, such that $J_Q = \sum_i Q_i^* J_i$. This means that a finite heat flux can still be maintained even at zero temperature gradient, as long as a nonzero ionic flux exists. $Q_i^*$ can therefore be interpreted as the heat carried along with the ionic flux of species $i$. The concept of "heat of transport" was proposed as early as 1926 by Eastman [60]. This quantity can be understood microscopically as shown in Fig. 2. Due to the temperature gradient, the local free energy density profile becomes non-uniform, and when an ion migrates from the region at temperature $T + dT$ to the region at temperature $T$, the associated free energy change is expressed as $[g_i(T) - g_i(T + dT)]v_i$, where $g_i$ is the local free energy density and $v_i$ can be viewed as the volume occupied by ion $i$. The derivative of local $g_i$ at nonequlibrium with respect to temperature can be viewed as the "entropy" carried with the ion transport, known as Eastman entropy of transfer:

$$s_i^* = \frac{[g_i(T) - g_i(T + dT)]}{dT} v_i = -\frac{\partial g_i}{\partial T} v_i. \tag{9}$$

with $-\partial g_i/\partial T$ the entropy density around the ion $i$. The entropy flux can be expressed as $J_S = J_Q/T$, thus Eastman entropy of transfer can be expressed as:

$$s_i^* = Q_i^*/T. \tag{10}$$

Clearly, such ionic migration along the temperature gradient would result in the exchange of solvent molecules with the surrounding, and the Eastman entropy of transfer would be determined by intermolecular forces. According to Eq. (9), the Eastman entropy of transfer can be understood as the entropy carried in the "effective" volume of the ion species $i$. Agar adopted this kind of picture and derived the expression Eastman entropy of transfer using a hydrodynamic method, by integrating the entropy density upon the flow field surrounding the ion due to ionic thermodiffusion [21]. When the ion is sufficiently small such that the perturbative flow field can be neglected, the Eastman entropy of transfer at the dilute limit is reduced to the solvation entropy, which can be qualitatively described by the Born model [61]:

$$s_i^* = -\frac{(z_i e)^2}{8\pi \epsilon r_i} \frac{d(\ln \epsilon)}{dT}. \tag{11}$$

where $z_i$ is the ionic valence, $\epsilon$ is the dielectric constant and $r_i$ is the hydrodynamic radius of the ion. However, the Born model only included the electric free energy and neglected the detailed microscopic configuration of solvent molecules in the ion field. As a result, the Born model could underestimate the thermopower of some nonaqueous electrolytes with large structure-making effects, such as octanol, dodecanol, and ethylene-glycol [22]. A more comprehensive model of the heat of transport $Q_i^*$ and $s_i^*$ was derived by Helfand and Kirkwood based on the Bogolyubov–Born–Green–Kirkwood–Yvon (BBGKY) hierarchy theory [62], highlighting the importance of interaction between ions and solvent molecules and microscopic configurations of solvents.

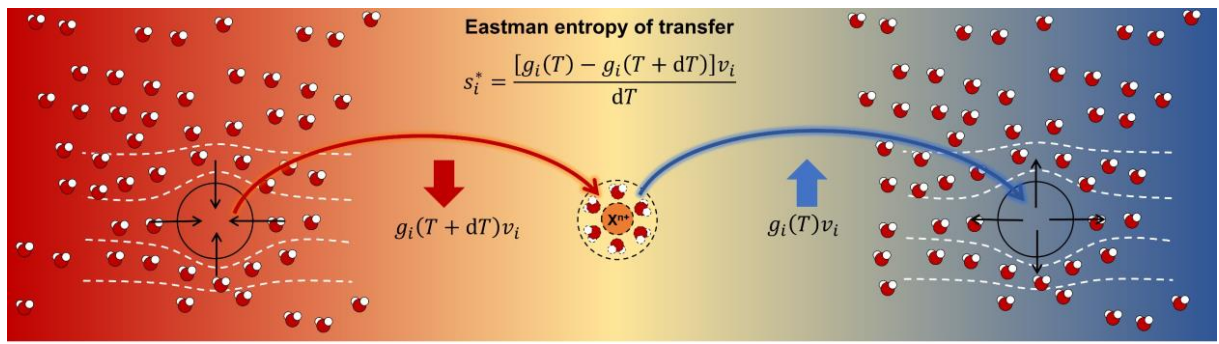

Figure 2. The physical picture of the Eastman entropy of transfer associated with the ionic thermodiffusion.

Although we have shown that the heat of transport $Q_i^*$ or equivalently the Eastman entropy of transfer $s_i^*$ is determined by the concentration gradient at nonequilibrium temperature, it is important to realize that the relation between Soret thermopower $S_{td}$ and the Eastman entropy transfer $s_i^*$ is also dependent on boundary conditions. Fig. 3a and Fig. 3b compare the difference in boundary conditions between a solid-state TEG device and a TIC device. In solid-state TEGs, electrons and holes can cross the interface between the thermoelectric material and the current collector, while the electrodes/current collectors in TIC devices are ion-blocking. Specifically, the open-circuit condition of the TEG device only requires the total current to be zero, while allowing ambipolar transport of electrons and holes; in contrast, the current of each ionic species should be zero for TIC devices. For TICs with ion-blocking electrodes, the thermopower is determined by the Eastman entropy of transfer but independent of ion mobilities:

$$S_{td} = \frac{s_+^* - s_-^*}{e(z_+ - z_-)}, \tag{12}$$

where the subscript $\pm$ denotes cations (+) and anions(-), $e$ is the elementary charge, $z_+ (> 0)$ and $z_- (< 0)$ are the valence of cations and anions, respectively. Fig. 3c. shows an open ionic channel connected to two reservoirs of electrolytes maintained at different temperatures, whose electrodes are permeable to ionic species. In this case, the open channel now allows ambipolar ionic current and the thermopower becomes mobility-dependent:

$$\tilde{S}_{td} = \frac{\mu_+ s_+^* - \mu_- s_-^*}{e(z_+\mu_+ - z_-\mu_-)}, \tag{13}$$

where $\mu_\pm$ the is ion mobility and the upper tilde (~) indicates that the system is open. The different behavior of thermopower in closed (TICs with ion-blocking electrodes) and open systems (ionic channel under temperature gradient) are quite different as shown in Fig 3d. Interestingly, $S_{td}$ and $\tilde{S}_{td}$ can even show opposite signs at a certain mobility ratio between the cation and anions. Eqs. (12-13) together pointed out the directions to search for high thermopower electrolytes. Thermopower of closed TIC devices can be enhanced by using electrolytes with the large mismatch of Eastman entropy of transfer instead of ionic mobilities, although there might be some correlation between the Eastman entropy of transfer and ionic mobilities. For ionic channels under a temperature difference, both the magnitude and the sign thermopower can be tuned by the mobility ratio between cations and anions. It is, however, important to note that Eqs. (12-13) are derived under a few approximations. (i) The Soret effect of the neutral solvent is completely ignored in this framework of Poisson-Nernst-Planck (PNP) theory. Therefore, Eqs. (12-13) are no longer applicable to the cases when the water content significantly affects the charge density profile. For example, the thermodiffusion of water in Nafion and S-PEEK membrane affects the hopping transport of proton and is an important factor contributing to thermopower in these membrane systems[13]. (ii) Eqs. (12-13) treat ions as point charges without volume such that the ion steric effects are not captured. Recent theoretical analysis showed that ion steric effects can be significant in concentrated electrolytes or ionic liquids under nanoconfinements[63]. (iii) Ionic correlations are ignored in PNP theory. When ion complexes are formed, the specific contribution to thermopower might not be consistent with the valence charge. For example, a positive ion can form negatively charged ion complexes and make a negative contribution to the total thermopower.

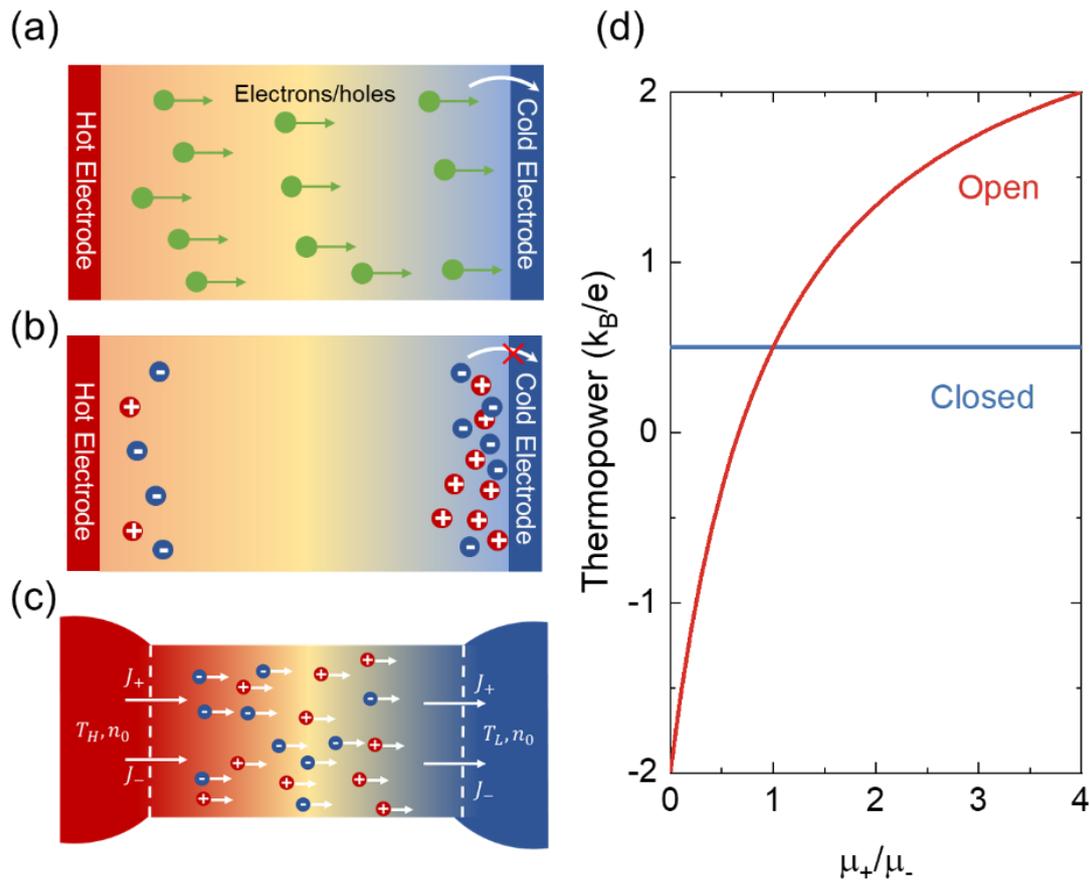

Figure 3. Thermoelectric effect of (a) holes/electrons in a semiconductor, (b) ions in a capacitor with ion-blocking electrodes, and (c) ions in an open channel connecting two reservoirs maintained under a temperature gradient. (d) Dependence of thermopower on mobility ratio for the open and closed i-TE devices.

**Strategies for enhancing Soret thermopower.** The past decade has seen significant progress in developing ionic thermoelectric materials with high thermopowers, such as nonaqueous solutions [22], nanoconfined electrolytes [11, 24-25], polyelectrolytes containing immobile poly-ions and mobile counterions [13, 23, 52], salts or ionic liquids in gel solvents [7, 12, 29-30, 52, 64], electrolytes containing complex ions [65], *etc*. This section analyzes these strategies in detail through the viewpoint of Eastman entropy of transfer.

*Structure making/breaking effect.* Since the Soret thermopower is closely related to solvation entropy for simple ions, tuning the solvation structure could be an effective way to enhance $S_{td}$. A notable example is the discovery of the huge Soret thermopower as large as 7 mV/K in nonaqueous Tetrabutylammoniumnitrate (TBAN) [22]. The authors discovered the square-root dependence of Soret thermopower on concentration, as predicted by Debye-Hückel theory [66]:

$$S_{td}(c) = S_{td}(0) + \frac{e^2 N_A}{4\pi\epsilon T}\kappa_D \left[\frac{1}{12} + \frac{3}{4}\frac{\partial \ln \epsilon}{\partial \ln T} - \frac{1}{4}\frac{\partial \ln \rho}{\partial \ln T}\right], \tag{14}$$

where $S_{td}(0)$ is the Soret thermopower at the dilute limit, $\epsilon$ is the dielectric constant, $\rho$ is the density and $N_A$ is the Avogadro number. However, by extrapolating experimental data with Eq. (13), $S_{td}(0)$ at the dilute limit reaches 8.8 mV/K for TBAN and 6.1 mV/K for tetradodecylammonium nitrate (TDAN) in dodecanol solvents (Fig. 4a), nearly two orders of magnitude larger than the prediction by Born solvation model (371 μK/V for TBAN). The Born model also severely underestimated the dielectric constant dependence of Soret thermopower, compared with experimental measurements of TBAN and TDAN in different solvents (Fig. 4b). Such discrepancy indicates that treating solvents simply as an effective dielectric medium fails to quantitatively predict the Soret thermopower. In fact, the existence of the ionic field would disturb the structure of the hydrogen bond network of the solvent. An ion can be a structure maker or breaker. The ionic field of a structure maker attracts solvent molecules to form a large solvation shell so that the solvent molecules become more structured than in the bulk solvent. In contrast, a structure-breaking ion tends to weaken the hydrogen network between the solvent molecules (Figs. 4c-d).

The structure-making/breaking effects can be quantified by the Gibbs free energy of transfer $\Delta G_{HB}$ proposed by Marcus and Ben-Naim [67], which enabled quantitative analysis of structure-making/breaking effects on $S_{td}$. A more positive (negative) $\Delta G_{HB}$ means a stronger structure making (breaking) effect. He et al. [30] performed a comprehensive study on tuning ionic thermopower using ions with different $\Delta G_{HB}$ in gel solvents. The authors found both cations and anions with stronger structure breaking effects tends to enhance $S_{td}$. With the same cation $K^+$ in PVA hydrogel, $S_{td}$ increases with increasing $-\Delta G_{HB}$ of different anions, which means that stronger structure breaking effects favor higher positive $S_{td}$. This is because structure breaking anions have thinner solvation shell and stronger interaction with hydroxyl groups in PVA, resulting in slower thermal drift along the temperature gradient. On the other hand, with the same anion I$^-$ in PVA hydrogel, a stronger structure-breaking cation also shows a more positive $S_{td}$. A structure breaking cation promotes the partial dissociation of hydroxyl groups (R-OH ⇌ R-O$^-$ + H$^+$), resulting in stronger repulsion of the anions. In turn, the concentration difference between cations and anions is increased, which leads to a higher Soret thermopower $S_{td}$.

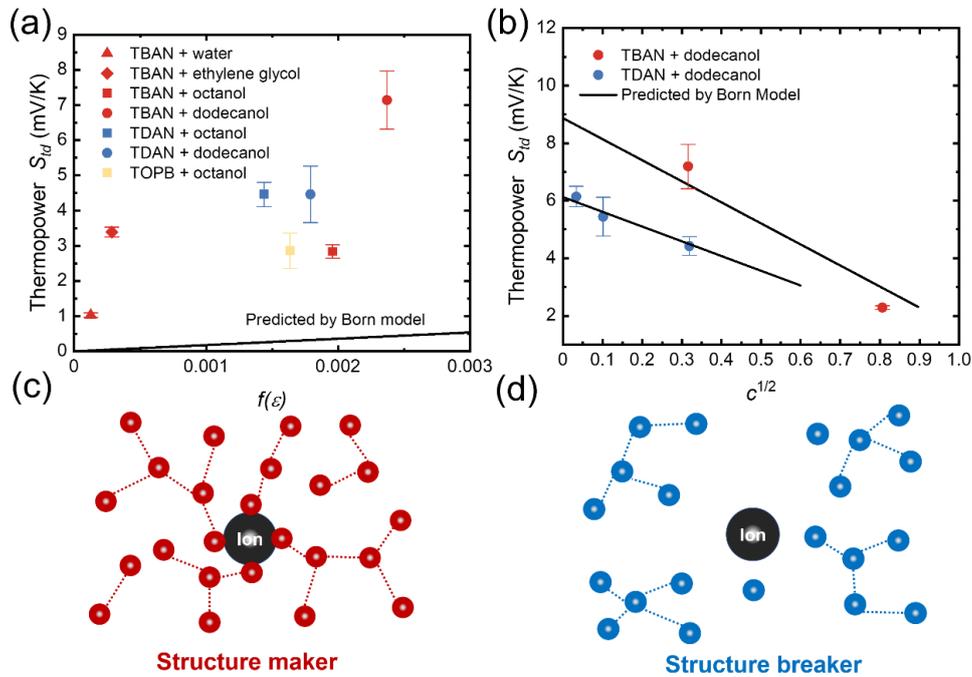

Figure 4. Structure making/breaking effects for Soret thermopower $S_{td}$ improvement. (a) Soret thermopower $S_{td}$ as a function of the square root of ion concentrations $c$ (mol/L) of TBAN and tetradodecylammonium nitrate (TDAN) in dodecanol. (b) Soret thermopower $S_{td}$ of TBAN salts in various solvents, as a function of dielectric constant $f(\epsilon) = -(r_i\epsilon^2)^{-1}d\epsilon/dT$. (c-d) The schematics of (c) the structure-making and (d) the structure-breaking effect. Readapted from ref. [22], Copyright 2011 with the permission of AIP publishing.

*Asymmetric cation-anion thermodiffusion.* In addition to structure making/breaking effect, leveraging asymmetries in $S^*$ of cations and anions is an effective way to enhance the thermopower. The large thermopower of 11 mV/K reported by Zhao *et al.* [23] in polyethylene oxide (PEO)-NaOH polyelectrolyte can be partially attributed to the fact that poly-ions are immobile so that the charge transport becomes much more unipolar. However, simply neglecting the thermodiffusion of poly-ions still cannot explain the experimentally observed large thermopower. Take proton conductive S-PEEK membranes for example. The Eastman entropy of transfer for protons in water is 44.3 J/(mol·K) [21], which only leads to an upper limit of Soret thermopower of 0.23 mV/K, still much lower than the measured ~ 5.5 mV/K [13]. A limitation of Agar's single-ion model is that it does not account for the specific configuration of solvent molecules in the vicinity of the ion. In polyelectrolytes and gel solutions, the polymer chains contain a large amount of charged or polar functional groups [7], whose interaction with the mobile ions makes the single ion picture breakdown. Such interaction could contribute significantly to the transported heat. For example, recent work by Huang's group also discovered that ion coordination between $Li^+$ and $BF_4$/TFSI anions in gel electrolytes can be used to bidirectionally tune the ionic thermopower [68], but detailed theoretical models fully capturing the ion correlation effects are still lacking.

Nanoconfinement could be another effective method to induce asymmetric thermodiffusion between cations and anions. When the distance between boundaries approaches Debye length, the electric double layer (EDL) becomes overlapped, resulting in the breakdown of local charge neutrality and selective transport of cations or anions. Dietzel and Hardt [25] theoretically proposed to improve thermopower by confining electrolytes in a slit open channel. Experimentally, Li et al. observed a huge thermopower of 24 mV/K of sodium polyethylene oxide (Na-PEO) electrolytes confined in cellulosic membranes (Fig. 5a) [11]. However, the experimental measurements for confined electrolytes are performed in a closed system with ion-insulating electrodes, whereas the theory by Dietzel and Hardt neglected the effects of ion-insulating boundaries[25]. It is therefore questionable whether the theory for long open channels can be used to interpret the experimental results. Qian et al. [24] developed an analytical model to describe the nanochannels confinement effect on Soret thermopower in both open and closed systems. For symmetrical electrolytes with equal valences $z = z_+ = z_-$, the thermopower of an open system $\tilde{S}_{td}$ and a closed system $S_{td}$ can be written as:

$$\tilde{S}_{td} = \frac{1}{e} \frac{s_+^* \mu_+ - s_-^* \mu_- - f_\psi (s_+^* \mu_+ + s_-^* \mu_-)}{\mu_+ + \mu_- - f_\psi (\mu_+ - \mu_-)}$$

$$S_{td} = \frac{1}{2e} [(s_+^* - s_-^*) - f_\psi (s_+^* + s_-^*)],$$

(15)

where $H$ is the gap of the nanochannel. The coefficient $f_\psi = \int_{-H/2}^{H/2} \sinh(z\Psi)\, dy / \int_{-H/2}^{H/2} \cosh(z\Psi)\, dy$ is determined by the dimensionless EDL potential $\Psi = e\psi/k_B T$, where $\psi$ denotes the electric potential in EDL. The sign of $f_\psi$ is the same as the sign of surface charges. As shown in Figs. 5b-c, a more negative surface potential $\psi_0$ and decreasing channel widths $H$ result in increasingly unipolar ion transport and an increased p-type thermopower. Interestingly, whether nanoconfinement can enhance the thermopower of polyelectrolytes strongly depends on whether the system is open or closed. Polyelectrolytes such as Na-PEO have intrinsically high thermopower ~11 mV/K [69], whose polymeric anions are almost immobile while the Na⁺ counterions are small and mobile. In this case, the mobilities satisfy $\mu_+ \gg \mu_-$. Since thermopower of an open channel depends on the mobilities, the thermopower for polyelectrolyte is dominated by counterions, resulting in $\tilde{S}_{td} \approx s_+^*/e$ independent of $f_\psi$. Therefore, nanoconfinement effects cannot improve the thermopower of electrolytes with extremely mismatched ionic mobilities in an open system. For closed systems, however, the overlapping effect of EDLs can effectively enhance thermopower, which is consistent with observations of large thermal voltage measured from the nanocellulose capacitors[11].

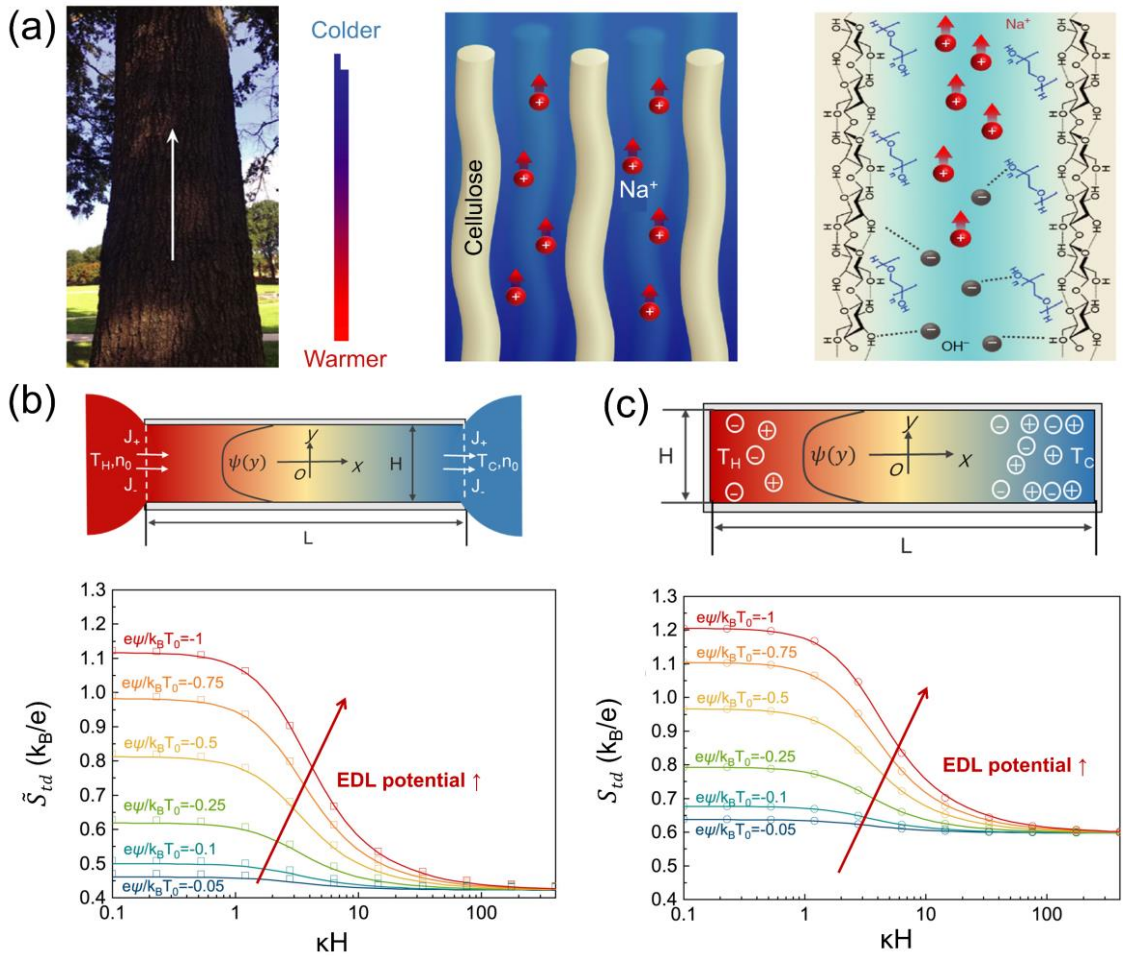

Figure 5. Nanoconfinement effect in channels. (a) Aligned cellulose nanofibers and the ion transport in cellulose nanochannels. From ref. [11], Copyright 2019, reproduced with permission from Springer Nature. Thermopower dependence on the potential of electric double layer (EDL) and channel widths of (b) closed capacitor and (c) open channel systems. Reprinted from ref. [24], Copyright 2022, with permission from Elsevier.

Page 20 / 76

*Effect of activation barrier.* In polyelectrolytes, gels, and ionic liquids, the coupling between charged species can be described by a hopping picture with an effective activation barrier (Figs. 6a-b). Specifically, the large amount of charged or polar groups in polyelectrolytes or gel networks resulted in multiple minima of potential energy surface (PES), where the ion transport is dominated by hopping between local minima of the PES. In ionic liquids, the diffusion and thermodiffusion of ions also obey the hopping dynamics according to the "hole theory" [70], which assumes that there exist vacancies created by density fluctuations. Driven by thermal fluctuations, ions could jump into the vacancies and leave behind a new vacancy. When a temperature gradient is established in ionic liquids, the ionic hopping rates become biased which leads to the thermodiffusion of ions. Such a hopping process requires ions to overcome the interaction energy with the surrounding ions, namely the activation enthalpy of hopping. Wüger[71] derived an expression of hopping thermopower $S_{td}$ using the Erying model [72]:

$$S_{td} = \frac{k_B}{e}(w_+ - w_-) + \frac{w_+ \Delta H_+ - w_- \Delta H_-}{eT}, \quad (16)$$

where the weighting coefficients are determined by the concentration of ions $w_\pm = n_\pm/(n_+ + n_-)$. The first term in Eq. (16) is the typical Soret thermopower, while the second term is the excess thermopower due to the hopping activation enthalpy, typically much larger than the Soret contribution. The thermopower of the ionic liquid EMIM-TFSI is predicted as -0.83 mV/K by this hopping theory (with $w_\pm$ and the hopping barriers $\Delta H_\pm$ derived from Hittorf transport numbers [73] and mobilities[52] respectively), agreeing well with experimentally measured thermopower -0.85 mV/K [52] (Fig. 6c). However, the activation barrier $\Delta H$ is still not enough to explain the huge thermopower observed in polyelectrolytes and gels such as KCl-gelatin (Fig. 6c). One possible reason for the theoretical underestimation is the assumption that compositions of species still remain uniform under non-uniform temperatures. The concentration profile of both ions and solvents in polyelectrolytes and gel could be sensitive to temperature due to the temperature-dependent ionization of polyions, thermosensitive coupling/decoupling between the mobile ions and the polymer backbones, and thermodiffusion of solvents. Such composition change could result in extra contributions to effective heat of transport or Eastman entropy of transfer, yet a theoretical model with predictive power is still lacking.

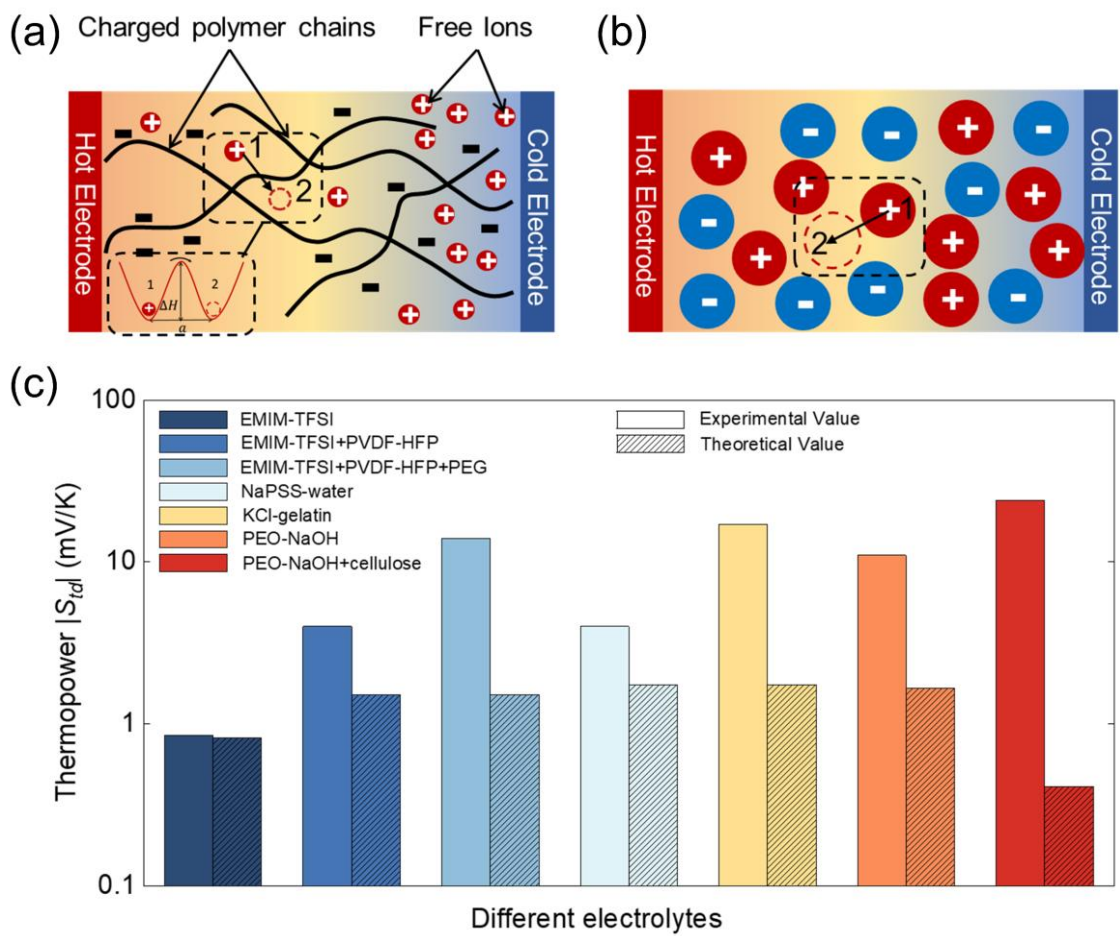

Figure 6. Ionic hopping and the Soret thermopower of polyelectrolytes and ionic liquids. (a-b) Ionic hopping in polyelectrolytes and ionic liquids. $\Delta H$ denotes the enthalpy barrier and $a$ is the average jump distance. (c) Experimental and theoretical thermopower in polyelectrolytes and ionic liquids, in which EMIM is 1-

ethyl-3-methylimidazolium, TFSI is bis(trifluoro-methylsulfonyl)imide, PVDF-HFP is poly(vinylidene fluoride-co-hexafluoropropylene), PEG is polyethylene-glycol, and NaPSS is polystyrene sulfonate sodium. Experimental data collected from ref. [71].

**The energy conversion efficiency of TIC devices.** The performance of traditional thermoelectric materials and devices is quantified using the thermoelectric factor $Z = S^2 \sigma k^{-1}$ or figure of merit $ZT$, which is determined by the conductivity $\sigma$, Seebeck coefficient $S$, and thermal conductivity $k$ [74-75]. Despite the similarity between TICs and thermoelectric generators, it is questionable to directly apply the figure of merit to evaluate the performances of TICs due to their capacitive nature in charging and discharging. [59, 76]. This section discusses the operation of TIC devices and proposes a modified thermoelectric factor $Z_{TIC}$ for proper evaluating performances of TICs, which is only well-defined on the device-level.

Figs. 7a-b show the four-stage operation mode of TIC devices [23, 77-78], using a p-type electrolyte as an example. First, a temperature difference is established across the capacitor, driving the thermodiffusion of ions and the generation of thermal voltage. Second, the two electrodes are connected to the external load $R_L$, and electrons would flow from the hot electrode to the cold electrode, which charges the electrode of TIC and outputs electric work. Third, the charged TIC is set to open-circuit again and relaxed to thermal equilibrium. Finally, electric work can be extracted again by discharging the TIC with an external load. The time consumed in these four stages is denoted as $t_{ch}$, $t_{dis1}$, $t_{eq}$, and $t_{dis2}$, correspondingly.

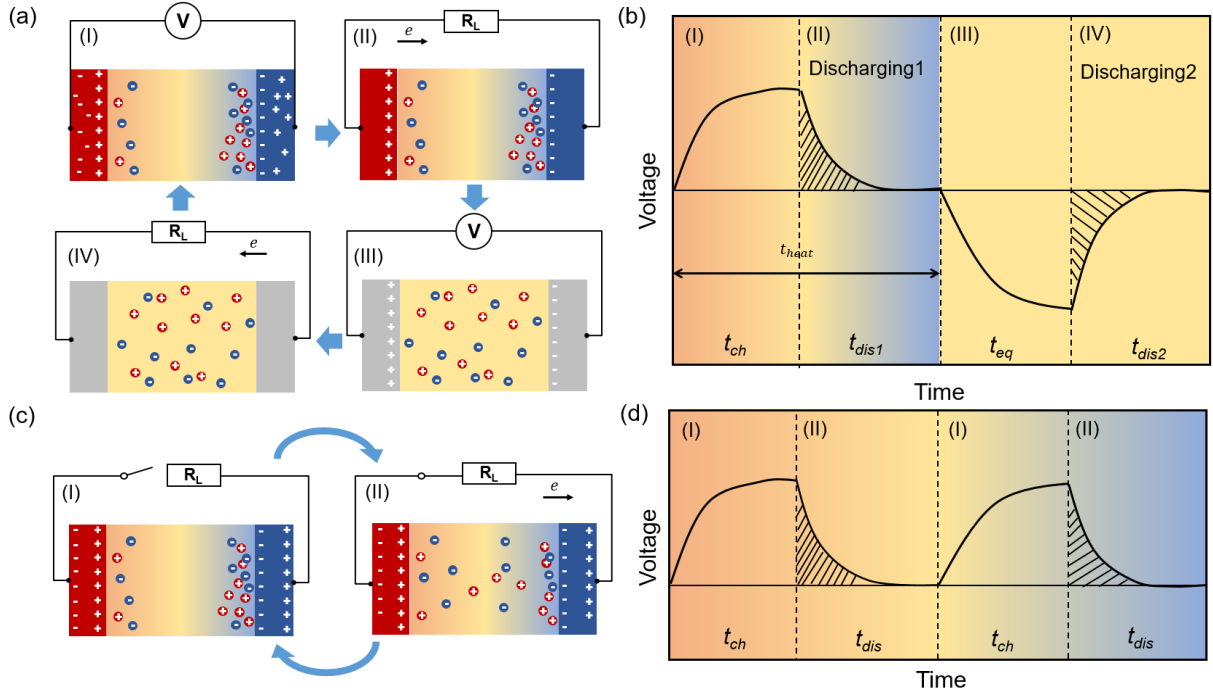

Figure 7. (a) Operation processes of TICs in the four-stage mode: (I) Thermal voltage is built up due to the thermodiffusion of ions when the temperature gradient is applied (II) discharging of the TIC by connecting the electrodes to an external load, (III) withdraw the heating to reach equilibration, and (IV) the second discharging. (b) The voltage of the whole cycle during the four stages: Apparently, the total heating time $t_{heat} = t_{ch} + t_{dis1}$. (c) Operation processes of TICs in the two-stage mode. Temperature gradient is always applied across the device. (I) Thermodiffusion of ions induces a thermal voltage across the device in an open circuit. (II) Discharging of TIC connected to an external load. The ion distribution is partially relaxed and the electrode is slightly polarized. (I) and (II) can be repeated hundreds of times until significant electrode polarization results in decreased voltage. The device can be easily regenerated by short-circuiting the electrodes at thermal equilibrium. (d) The voltage of the thermal charging/discharging two-stage mode.

The efficiency of TIC devices is defined as the ratio of power generation and heat absorption:

$$\eta_E = \frac{W}{Q_P + Q_F - Q_J}, \quad (17)$$

where $W$ is the electric work, $Q_P$ is the Peltier heat, $Q_F$ is the heat conduction across the TIC, and $Q_J$ is the Joule heat. For the four-stage operation mode, the total electric work is calculated as $W = C(S_{td}\Delta T')^2$, where $C$ is the capacitance, and $\Delta T'$ is the temperature difference across the device, which is usually smaller than the temperature difference $\Delta T$ between the two reservoirs taking into account the heat transfer process. Now we articulate that charging and discharging timescales are important for enhancing efficiency. Both the Peltier heat and Joule heating are determined by the discharging time, calculated as $Q_P = S_{td}T_H \left( \int_0^{t_{dis1}} I dt + \int_0^{t_{dis2}} I dt \right)$, and $Q_J = \int_0^{t_{dis1}} I^2 R_L dt + \int_0^{t_{dis2}} I^2 R_L dt$, where $I(t) = I_0 \exp\left(-\frac{t}{R_L C}\right)$ is the transient current of the capacitor. The heat conduction loss is $Q_F = kA\Delta T' L^{-1} \sum_j t_j$, with $\sum_j t_j$ denoting the total time consumed by the four charge-discharge stages. The discharging time can be approximated as three times the time constant of the R-C circuit to ensure a 95% discharge level, expressed as $t_{dis1} = t_{dis2} = 3R_L C$. Reducing the thermal charging time $t_{ch}$ is the key to improving efficiency. The timescale $t_{ch}$ can be approximated as $t_{ch} = \max(t_t, t_i)$, where $t_i = L^2/D$ is the characteristic time for ion diffusion, and $t_t$ is the timescale for establishing the temperature field. $t_t$ can be expressed as $3L^2/[af(\text{Bi})]$, where $L$ is the device length, $a$ is the thermal diffusivity, Bi is the Biot number quantifying the coupling with the heat reservoir and $f(\text{Bi})$ is a general eigenfunction increasing monotonically with Bi. $f(\text{Bi})$ is obtained by solving the transient heat transfer equations using the separation of variables and taking the leading term of the series expansion of the solution. For convenience of computation, we first numerically compute $f(\text{Bi})$ and obtained an approximated expression: $f(\text{Bi}) \approx 9.79 - \left(3.16 e^{-\text{Bi}/31.84} + 6.61 e^{-\text{Bi}/3.84}\right)$ (see Supporting Information S1).

Substituting these timescales into Eq. (16), the efficiency of the four-stage operation can be expressed as (see Supporting Information S2):

$$\eta_{E-4s} = \frac{Z_{TIC} T_H}{2 Z_{TIC} T_H + 1} \cdot \frac{\eta_C}{1 + 2/\text{Bi}}, \tag{18}$$

where $\eta_C = 1 - T_C/T_H$ is the Carnot efficiency, and the $Z_{TIC}$ factor is defined as:

$$Z_{TIC} = \frac{CS_{td}^2}{K(t_{ch} + 3R_L C)}, \tag{19}$$

where $K$ is the thermal conductance of the TIC device. Eqs (18-19) explicitly demonstrate the importance of reducing the thermal charging time for optimizing the efficiency of TIC devices with four-stage operation modes. Reminded that $t_{ch}$ for the four-stage mode is approximated as $t_{ch} = \max(3L^2/[af(\text{Bi})], L^2/D)$. Therefore, the effect of heat coupling with the reservoir, together with the specific heat of the system is already included in this derivation. Eq. (18) indicates that the theoretical limit of efficiency for TIC devices is half of the Carnot efficiency. Recently, Song et al. [76] proposed a thermoelectric factor $Z'_{TIC} = \frac{CS_{td}^2 G}{k\rho c_p}$ to evaluate TIC devices with a four-stage working mode, where the interface conductance $G$ between TIC and the heat reservoir is considered for the first time. Nevertheless, their model is no longer valid at very large Bi, and the thermoelectric factor $Z'_{TIC}$ diverges at the limit $\text{Bi} \to \infty$. The divergence of $Z'_{TIC}$ means that, the efficiency is always equal to the theoretical maximum $\eta_C/2$ under the fixed temperature boundary condition, which is unphysical. We suggest using Eq. (18) for thermodynamic analysis of TICs, which is valid both at fixed temperature conditions and under finite thermal conductance with the reservoirs.

Fig. 7c-d shows the two-stage operation of TICs, which is adopted by Han et al[7]. Different from the four-stage operation, the temperature gradient is always applied across the TIC device. The thermal charging process is similar to the four-stage operation, where the thermodiffusion of ions induces a thermal voltage. During the discharging process, electrons flow from the external circuit, and the ions inside the TIC redistribute resulting in a partially relaxed concentration profile. The electrodes become slightly polarized after the discharging process

(Fig. 7c). Then the thermal charging/discharging processes can be repeated hundreds of times under the steady-state temperature gradient until the thermal voltage decreases and electrodes are polarized significantly. The TIC can be easily recovered by short-circuiting the two electrodes at thermal equilibrium. Since the time consumed by the regeneration is typically negligible compared with the repeated thermal charging/discharging stages, we can neglect the regeneration process in thermodynamic analysis[7]. Following the similar procedure of modeling ion transport in the four-stage mode, the efficiency of the two-stage operation mode can be expressed as:

$$\eta_{E-2s} = \frac{Z_{TIC}T_H}{2(Z_{TIC}T_H + 1)} \cdot \frac{\eta_C}{1 + 2/\text{Bi}}, \tag{20}$$

Detailed derivation can be found in Supporting Information. Figs. 8a-b presents the influence of Biot number Bi and diffusion coefficient $D$ on the relative efficiencies defined as $\eta_{r-i} = \eta_{E-i}/\eta_C$, where $i = 2s$ or $4s$ denoting the two-stage or four-stage operations modes, respectively. For both the four-stage and the two-stage operations, increasing interface conductance with the reservoirs helps to enhance efficiency by suppressing the interfacial temperature drops. In general, the four-stage mode has higher efficiency compared with the two-stage mode, because the temperature gradient is always applied in the two-stage operation, resulting in higher heat conduction losses.

With the efficiency and the thermoelectric factor derived, one can evaluate how different factors can affect the performance of TICs. For example, the hopping activation barrier $\Delta H$ discussed before can contribute significantly to the thermopower $S_{td}$, but higher barriers are usually detrimental to ion transport. According to the Einstein–Smoluchowski theory [79] of Brownian motion, the diffusion coefficient scales exponentially with the hopping barrier: $D =$

$D^* \exp(-\Delta H/k_B T)$. Based on Eqs. (19-20), efficiency change with respected to the hopping barrier can be predicted. Fig. 8c shows the coupling between $S_{td}$ with the ionic hopping barrier [80]. Pursuing high $S_{td}$ is not always beneficial for maximizing the efficiency when the hopping barrier is larger than an optimal value (the same order as $k_B T$), since an overlarge barrier would suppress the ionic diffusivity.

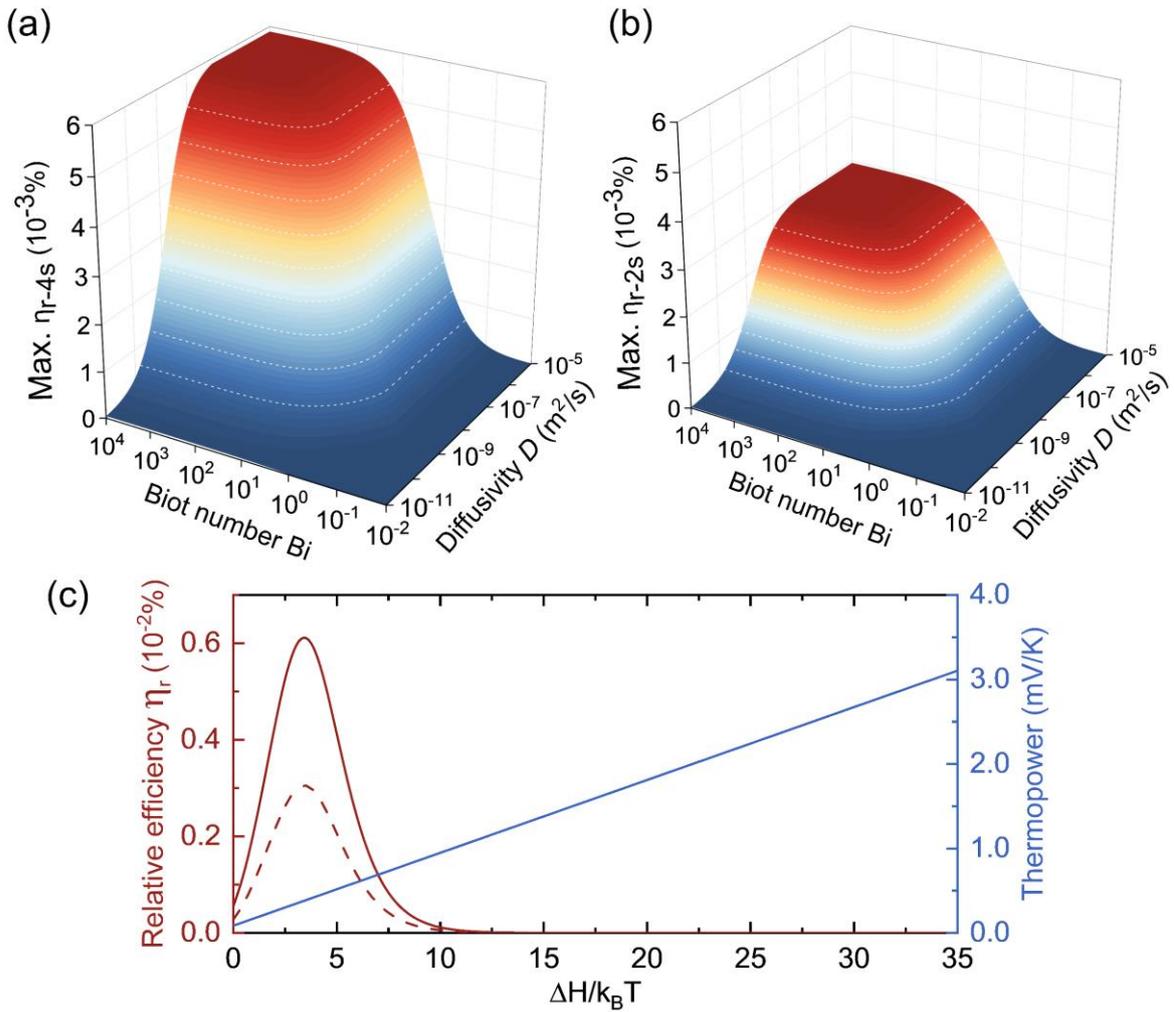

Figure 8. Relative efficiency of (a) the four-stage operation $\eta_{r-4s}$ and (b) the two-stage operation $\eta_{r-2s}$ versus Biot number $Bi$ and diffusion coefficient $D$. (c) The relative efficiency and thermopower affected by the activation barrier $\Delta H/k_B T$.

**Electrode engineering and device design of TICs.** In addition to electrolyte properties such as the Soret thermopower $S_{td}$ and the ion diffusivity and electrode-electrolyte interfaces are also significant in determining the performances of TICs. First, electrodes can contribute to an extra thermal voltage of TICs, due to the thermal sensitivity induced by the temperature-dependent capacitance or the Donnan potential at the electrode surface [81]. In addition, the morphology of the electrodes determines the capacitance $C$ of TICs, which is significant in improving energy density and efficiency. Finally, structure of TICs can be engineered for shortening thermal charging time using a stacking electrode design.

*Thermal sensitivity of electrodes.* In addition to the Soret effect of electrolytes, Bonetti *et al.* [82] pointed out that the thermal sensitivity of electrodes contributes to the effective thermopower $S_e$, such that $S_e = S_{td} + d\phi/dT$ where $d\phi/dT$ is the derivative of interface electrostatic potential $\phi$ to temperature, namely the thermal sensitivity of electrodes. Electrode materials are typically metallic due to their high electrochemical activity, capacitance, and conductivity [83-85]. The thermal sensitivity of the metallic electrode is correlated to the work function of the electrode, defined as the minimum energy required for an electron to escape from the metal to vacuum [86]. The mechanism of the thermal sensitivity can be explained by the Jellium model of the electric double layer [87]. The capacitance of the electric double layer at the interface consists of the capacitance of the Helmholtz layer $C_H$, the diffusion layer $C_D$, and the capacitance originated from the electron spillover on the electrode surface $C_M$ (Fig 9a). The thermal sensitivity is attributed to the temperature-dependent $C_M$ as $\frac{d\phi}{dT} = \frac{1}{C_M}\left(\frac{dQ}{dT} - \phi_M \frac{dC_M}{dT}\right)$, where $Q$ is the effective surface charge and $\phi_M$ is the interface potential at the electrode surface. According to the Jellium model, the $C_M$ is dependent on the density of positive ions in the electrode $n_+$ ($C_M^{-1} \propto n_+$). Lim *et al.* [85] characterized the thermal sensitivity of metal electrodes, including indium, copper, nickel, and platinum, and discovered that metals with a high work function tend to have a higher thermal sensitivity, with platinum showing a high thermal sensitivity of 6.5 mV/K (Fig. 9b). Different from metal electrodes, thermal sensitivity of polymer electrodes can be explained by the Donnan exclusion effect. As shown in Fig. 9c, the charged functional groups in the poly-electrode can generate a temperature-dependent electrostatic field, as expressed by the ratio between counterion concentrations inside and outside the polymer electrode: $\frac{d\phi}{dT} = \frac{\mathcal{R}}{nF} \ln \frac{c_{in}}{c_{out}}$, with $\mathcal{R}$ the ideal gas constant. Mardi et al. [88] measured the output potentials of the TIC device with electrodes that have different contents of polystyrene sulfonate (PSS) in electrolytes of ionic liquid polymer gel EMIM-TSFI+PVDF-HFP, with the optimized electrode showing a thermopower of -4.58 mV/K (Fig. 9d).

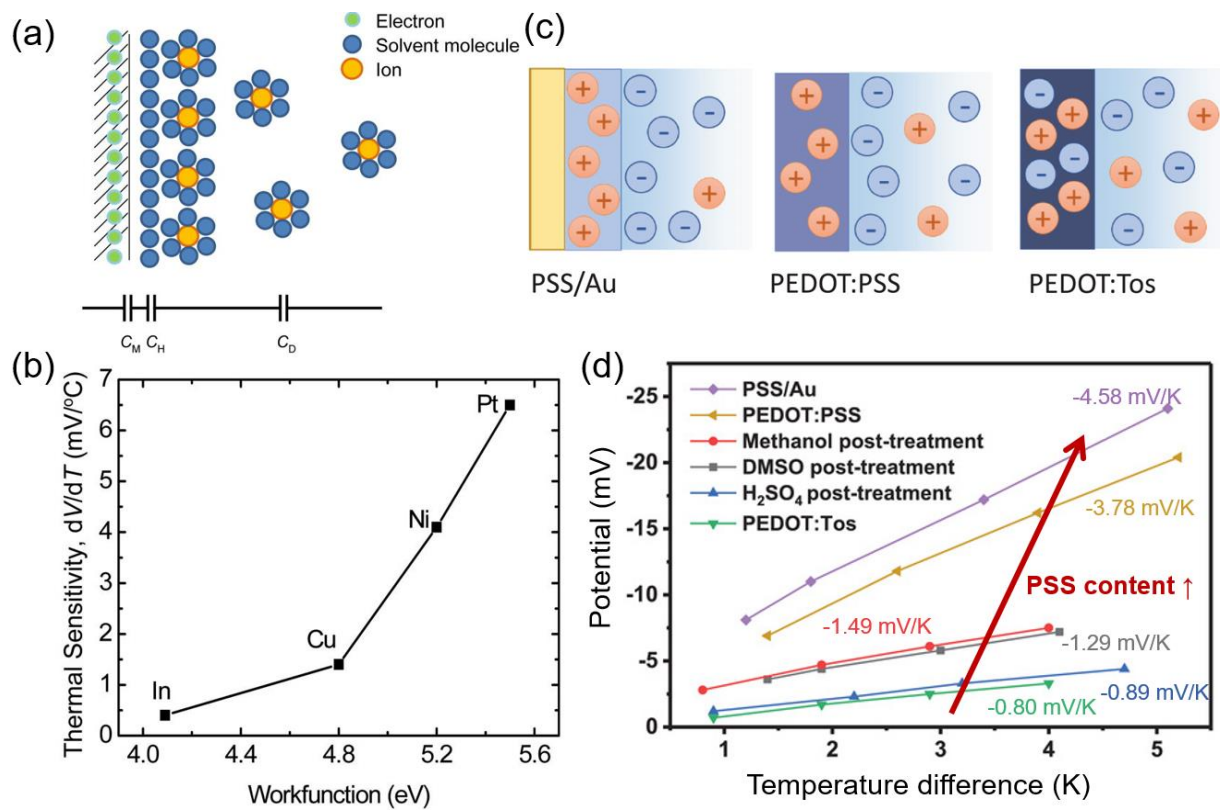

Figure 9. Thermal sensitivity of electrodes in TICs. (a) Schematic of the effective capacitances in the electric double layer. (b) The relationship between the thermal sensitivity and the work function. Reprinted from ref. [85], Copyright 2015 with the permission of AIP Publishing. (c) Illustration of the charge distribution at the electrode/electrolyte interface. Readapted from ref. [88], Copyright 2021 by Wiley. (d) Thermopower of different electrodes, with data collected from refs. [85, 88].

*Thermal charging time.* Nanostructured and porous electrodes [23, 27, 52, 77] can effectively improve the capacitance of TICs due to the high specific surface area [89-90]. However, multiscale porous structuring of the electrodes could also result in slow charging processes with characteristic time orders of magnitude larger than the ion diffusion time $1/D\kappa_D^2$ across the EDL, with $D$ and $\kappa_D$ the ion diffusivity and inverse Debye length. This slowing down of charging processes has been observed in porous electrodes in supercapacitors [91]. Since the ion transport in the charging process of TICs is similar to that of supercapacitors, slow charging could also lead to low efficiency when the TIC is operated in the two-stage mode. For the 4-stage mode, the thermal charging time is dictated by the longer one of the time scales for establishing the temperature gradient and the local charge density. One possible method of mitigating the sluggish thermodiffusion of ions is by adding extra electrodes (Fig. 10a), which is equivalent to connecting multiple thin capacitors in series (Fig. 10b). This kind of stacking electrode configuration would not affect the total capacitance, but the thermal charge time can be decreased with a shorter distance of ionic thermodiffusion (see Supporting Information S3). Fig. 10c shows that the efficiency can be improved by stacking electrodes in TICs, until the heat absorption during the thermal charging time cannot be further reduced through electrode stacking.

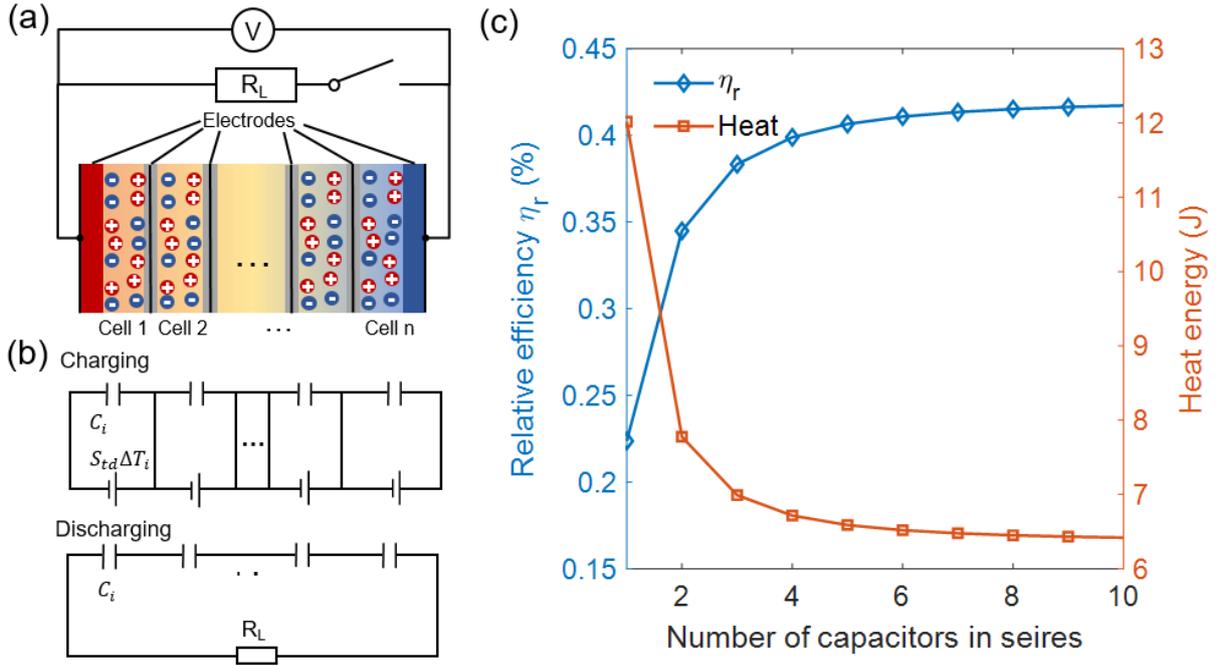

Figure 10. (a) A schematic of TIC stack containing $n$ sub-capacitor cells. (b) The equivalent circuit when charging and discharging. (c) Predicted heat absorption and relative efficiency versus the cell number.

- **THERMOGALVANIC CELLS**

TGCs leverage the temperature dependence of electrochemical reactions to convert the temperature difference to electricity. Compared with TIC devices, the major advantage of TGCs is the capability of continuous power output due to the existence of Faradaic redox reactions, although the effective thermopower of TGCs is typically smaller than that of TICs. First, we summarize recent advancements in improving $S_{tg}$. Driven by the voltage generated by the thermogalvanic effect, the electric currents are generated through Faradaic redox reactions at the electrode-electrolyte interfaces. Tailoring interfacial properties and the reaction kinetics are the key to enhancing the efficiency and power density of TGCs. However, current literature largely overlooked these interface effects and directly used the figure of merit $ZT$ based on electrolyte properties for evaluating efficiencies. We then discuss the impact of interface reactions on the TGCs' efficiency and propose a modified figure of merit $Z_{TGC}$, which incorporates the effect of ion transport and electrochemical reaction kinetics. The effects of electrode engineering and device configurations on power density and efficiency are discussed.

**Strategies for enhancing thermogalvanic thermopower.** The thermogalvanic thermopower $S_{tg}$ quantifies the temperature dependence of electrochemical potential $E$, and we show that the thermopower $S_{tg}$ is essentially related to the reaction entropy change. The electrochemical potential $E$ of a generic redox reaction $O + ne \rightarrow R$ is related to the free energy through $E = -\Delta G_{rxn}/nF$, where $\Delta G_{rxn} = G_R - G_O$ is the molar free energy change when the oxidized state $O$ is reduced to the reduced state $R$, $n$ is the number of electrons transferred and $F$ is the Faraday constant. The thermogalvanic thermopower $S_{tg}$ can then be expressed as:

$$S_{tg} = -\frac{E(T+dT)-E(T)}{dT} = \frac{1}{nF}\frac{d\Delta G_{rxn}}{dT} = -\frac{\Delta s_{rxn}}{nF}. \tag{21}$$

$\Delta S_{rxn}$ is the reaction entropy change:

$$\Delta s_{rxn} = -\frac{d\Delta G_{rxn}}{dT} = s_R - s_O, \tag{22}$$

where $s_R$ and $s_O$ are the partial molar entropy of the reduced and oxidized state respectively. Eqs. (21) indicate that increasing the reaction entropy change $\Delta s_{rxn}$ is the key to enhancing the thermogalvanic thermopower $S_{tg}$.

To further clarify possible mechanisms contributing to $\Delta S_{rxn}$ and $S_{tg}$, we can refer to the Nernst equation of electrochemical potential $E(T) = E^0(T) + (\mathcal{R}T/nF)\ln(a_O^{\nu_O}/a_R^{\nu_R})$, with $\mathcal{R}$ the ideal gas constant, $a_{O/R}$ the activity, and $\nu_{O/R}$ the stochiometric number. By taking the derivative with respect to temperature, the thermogalvanic thermopower can be written as:

$$S_{tg} = -\frac{1}{nF}(\Delta s_{rxn}^0 + \Delta s_{Nernst} + \Delta s_a), \tag{23}$$

where $\Delta s_{rxn}^0$ is the standard reaction entropy change at unit activity ratio. $\Delta s_{Nernst}$ is the Nernst contribution to reaction entropy and $\Delta s_a$ is the entropy originated from the nonuniform activity, expressed as:

$$\Delta s_{Nernst} = \mathcal{R} \ln \frac{a_O^{v_O}}{a_R^{v_R}} \tag{24}$$

$$\Delta s_a = \mathcal{R}T \frac{\mathrm{d}}{\mathrm{d}T} \ln \frac{a_O^{v_O}}{a_R^{v_R}} \tag{25}$$

Eq. (23) points out the possible strategies for enhancing $S_{tg}$, including (1) developing redox electrolytes with intrinsically high standard reaction entropy change $\Delta s_{rxn}^0$; (2) tuning the concentration ratio $C_O/C_R$ to enhance the Nernst contribution $\Delta s_{Nernst}$; (3) introducing thermosensitive mechanisms to maintain a large gradient of concentration/activity ratio, such that $\Delta S_{dact}$ can be enhanced. However, a nonstandard concentration (activity) ratio typically results in a small limiting current at either the cathode or the anode and thereby a suppressed power output, therefore strategy (2) is not feasible. This section will review recent advances in tuning reaction entropy change and leveraging thermosensitive activity ratio for developing high thermopower TGCs.

*Tuning reaction entropy change.* The redox entropy change consists of a variety of mechanisms, including the configurational entropy $\Delta s_{conf}$, the phonon entropy $\Delta s_{ph}$, the electron entropy $\Delta s_e$ and the solvation entropy $\Delta s_{solv}$. In most cases, $\Delta s_e$ is negligible unless metal-insulator phase transition is involved in the redox reaction [92]. The configuration entropy originates from the changes in atomic configurations associated with the redox reaction and is typically important in intercalation electrodes such as $Li_xCoO_2$ and $Li_xFePO_4$ [92-94]. The phonon entropy is due to the vibrational spectrum change after the redox reaction of the electrode material. For example, Gao et al. [95] observed a larger $S_{tg}$ of hexacynoferrate electrodes when the magnitude of lattice parameter change is significant after the sodium ion interaction, and they showed that the phonon entropy $\Delta s_{ph}$ originated from the vibrational spectrum change is mainly responsible. The solvation entropy $\Delta s_{solv}$ is related to the solvation structure change when ionic species in liquid electrolytes participate in the redox reaction. When both the oxidized and the reduced species are in the solution phase while chemical species from the electrode are not reduced or oxidized in the electrochemical reaction, the solvation entropy is the dominant contributor to $S_{tg}$. For TGCs based on simple liquid redox pairs such as $Fe^{3+}/Fe^{2+}$ and $Fe(CN)_6^{3-}/Fe(CN)_6^{4-}$, the reaction entropy is dominated by solvation entropy, so tuning $\Delta s_{solv}$ is the key to improving $S_{tg}$. Recent experiments showed that $\Delta s_{solv}$ can be enhanced by introducing nonaqueous solvents. For instance, Kim et al. [96] and Liu et al. [97] demonstrated that the addition of organic solvents to aqueous $Fe(CN)_6^{3-}/Fe(CN)_6^{4-}$ and $Fe^{3+}/Fe^{2+}$ electrolytes can lead to high $S_{tg}$ values of 2.9 mV/K and -2.5 mV/K, respectively. Inoue et al. [40] systematically studied the solvent effect on the thermopower of $FeCl_3/FeCl_2$ and found that $Fe^{3+}/Fe^{2+}$ in acetone solvents can reach 3.6 mV/K, nearly three times the thermopower of the pristine aqueous solution. This solvent dependence of thermopower of $Fe^{3+}/Fe^{2+}$ is related to the $D_{4h}$-type deformation of the $FeL_6$ (L denotes ligands) octahedra. Chen et al. [98] recently provided a molecular-level insight into the solvent effect on $S_{tg}$. Using molecular dynamics simulations (Fig. 11a), the $S_{tg}$ of redox pair $Fe^{3+}/Fe^{2+}$ is predicted through the free energy perturbation method. The authors further conducted comprehensive analysis on solvation structure and found that the larger difference in probability distribution of solvent dipole orientations would result in a larger $S_{tg}$ (Figs. 11b-c). This work showed that change in the solvation order after the redox reaction is the key to increase $S_{tg}$.

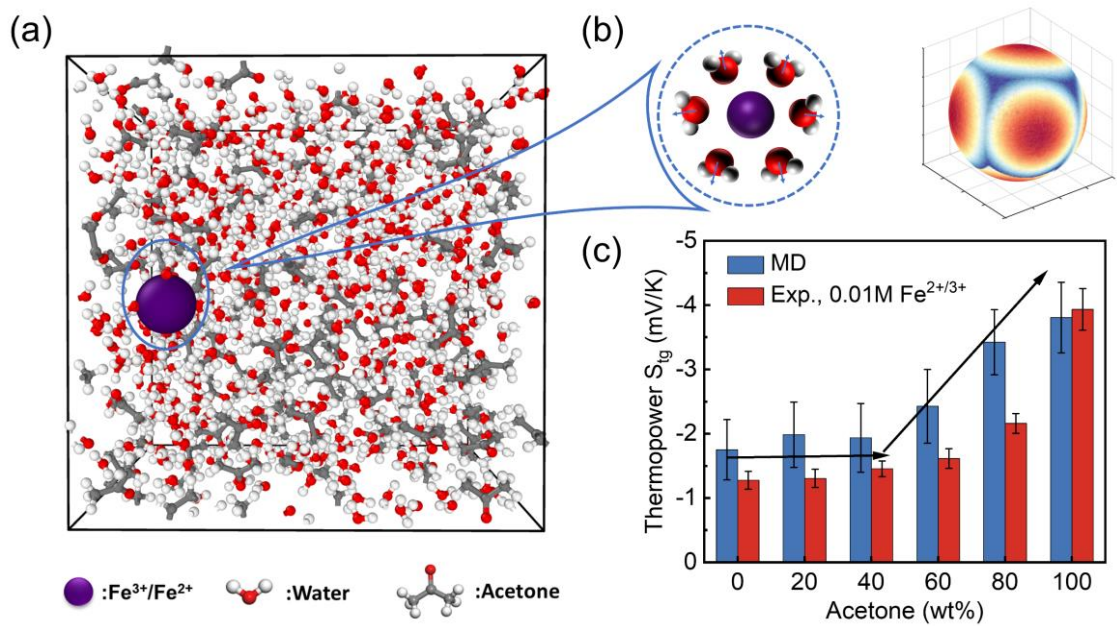

Figure 11. Solvent effect on the $S_{tg}$. (a) Snapshot of acetone molecules and $Fe^{3+}/Fe^{2+}$ ion in the simulation box. (b) Schematic of the first solvation shell around $Fe^{3+}/Fe^{2+}$ ion (left) and the trajectory of dipole orientations of water molecules in the first solvation shell of hydrated Fe ions (right). (c) Dependence of $S_{tg}$ on the acetone weighting fraction $\phi_{wt}$ by MD and by measurements. Reprinted from ref. [98], under the terms of the Creative Commons Attribution License, Copyright 2023 with permission by Wiley.

Another effective way to improve thermopower is by introducing electrolyte additives. Huang et al. [41] found that the addition of redox-inert supporting electrolytes can effectively enhance $\Delta s_{solv}$ and thereby $S_{tg}$ due to the structural change in the hydrogen bond network between water molecules. They attributed this effect to the structure-making and structure-breaking effects of supporting ions (Figs. 12a). A structure-making ion can enhance the proton network and attract water molecules to form a large solvation shell [99], where water molecules are more structured than those in pure bulk water. Oppositely, a structure-breaking ion interacts weakly with water molecules, making solvent molecules in the solvation shell of redox ions more disordered. Fig. 12b shows the linear correlation between the change in reaction entropy and the structural entropy solvents after adding supporting ions.

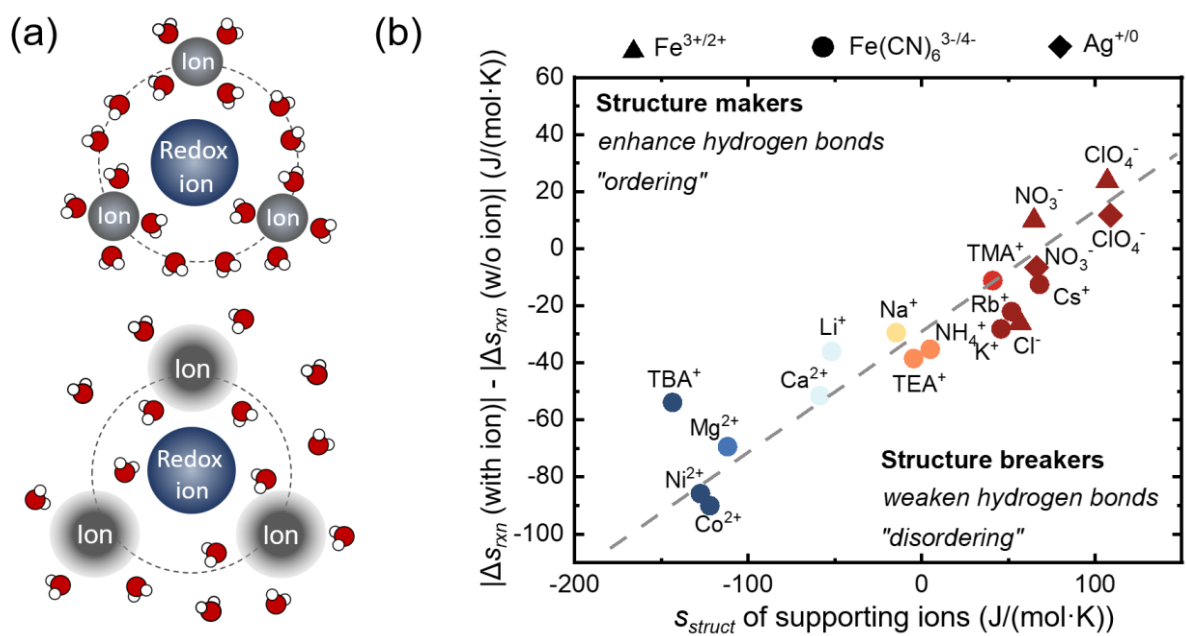

Figure 12. Effect of supporting electrolytes on the $S_{tg}$. (a) Solvation shell structure order changes by adding structural making (upper) and breaking ions (lower). (b) Change of redox reaction entropy of hydrated $Fe^{3+}/Fe^{2+}$ after adding supporting ions. Readapted from ref. [41] under the Creative Commons Attribution 3.0 Unported Licence, Copyright 2018 by RSC.

*Leveraging thermosensitive activity ratio.* According to Eqs. (23) and (25), establishing temperature-dependent activity or concentration ratio between the hot side and the cold side could contribute to excess $S_{tg}$, which can be achieved by leveraging thermosensitive chemistry. For example, Zhou *et al.* [100] discovered that adding α-cyclodextrin (α-CD) into $I_3^-/I^-$ aqueous electrolyte can induce thermosensitive encapsulation of the $I_3^-$ ion at the cold electrode, leading to an increased $S_{tg}$ from -0.86 mV/K to -1.97 mV/K. (Figs. 13a-b). The recent breakthrough in the power density and efficiency of TGC is made by Yu *et al.* [42] by introducing guanidinium ions (Gdm+) to selectively induce $Fe(CN)_6^{4-}$ thermosensitive crystallization to improve the activity difference between the hot and cold sides. In this TGC cell, the cold anode and the hot cathode are arranged vertically with the cold electrode above the hot electrode. On the cold side, $Fe(CN)_6^{3-}$ is reduced to $Fe(CN)_6^{4-}$ and crystalizes with $Gdm^+$. Then the crystals precipitate to the hot electrode driven by gravity, and $Fe(CN)_6^{4-}$ is dissolved again into the liquid phase and got oxidized at the hot anode (Figs. 13c-d). This single-ion thermosensitive crystallization of $Fe(CN)_6^{4-}$ contribute to excess activity difference across the device, and a high p-type $S_{tg} = 3.7$ mV/K is measured.

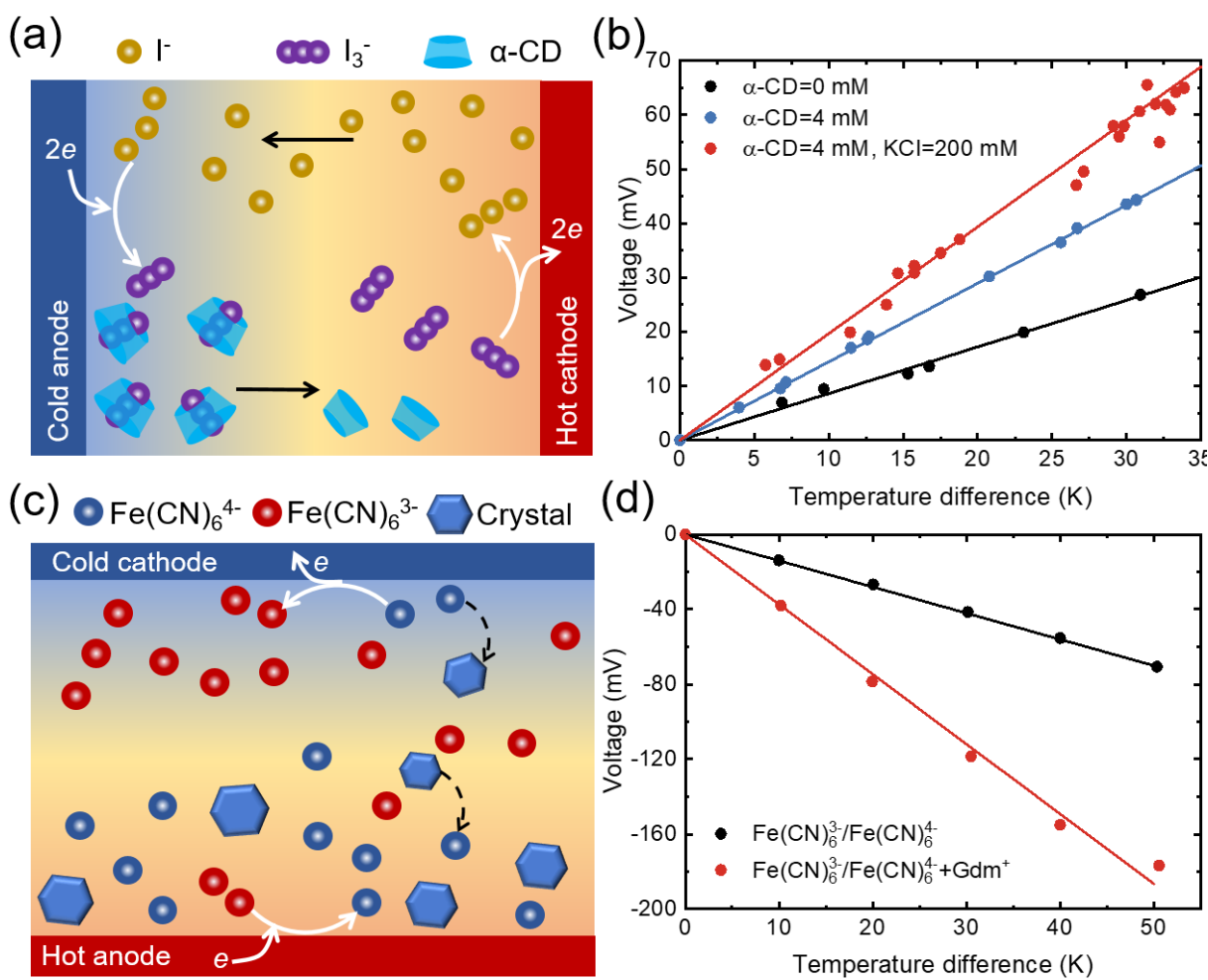

Figure 13. Inducing concentration gradients of redox ions for improved $S_{tg}$. (a) Concentration gradients established by introducing α-CD as thermosensitive acceptors of the $I_3^-$ ion, and (b) Thermopower improvement by adding α-CD. Readapted from ref.[100], Copyright 2016 by ACS. (c) Thermosensitive crystallization of $Fe(CN)_6^{4-}$ with guanidinium cations, and (d) Thermopower enhancement by adding guanidinium cations. From ref. [42], Copyright 2020, reprinted with permission from AAAS.

**Efficiency and figure of merit of TGC devices.** The efficiency of traditional thermoelectrics is determined by the figure of merit $ZT$. Nevertheless, there has been no consensus on the method of evaluating the energy-conversion efficiency of thermogalvanic cells. By simple analogy with thermoelectric generators, the $ZT = S^2\sigma T/k$ is usually adopted to evaluate the efficiency of TGCs. However, some studies directly used the intrinsic properties of electrolytes. For example, Abraham et al. [49] proposed to use the diffusivity of the redox ion dictating the limiting current when estimating $ZT$, yet other works used experimental values from electrochemical impedance spectroscopy (EIS) or the discharging voltage-current relations. An important concern for just using $ZT$ or $Z$ factor based on electrolyte properties for evaluating efficiency is the neglection of electrode-electrolyte interfaces where the redox reactions take place and electron transfer occurs. At finite discharging current, activation or mass transfer overpotential could significantly affect the voltage-current curves and the harvested power (Figs. 14 a-b). These overpotential losses near the electrode-electrolyte interface can account for a power density drop of over 50% compared with an ideal nonpolarizable electrode (Figs. 14 c-d). Incorporating these overpotential losses in thermodynamic analysis is clearly important to accurate evaluation of conversion efficiency, but is usually neglected by current literature.

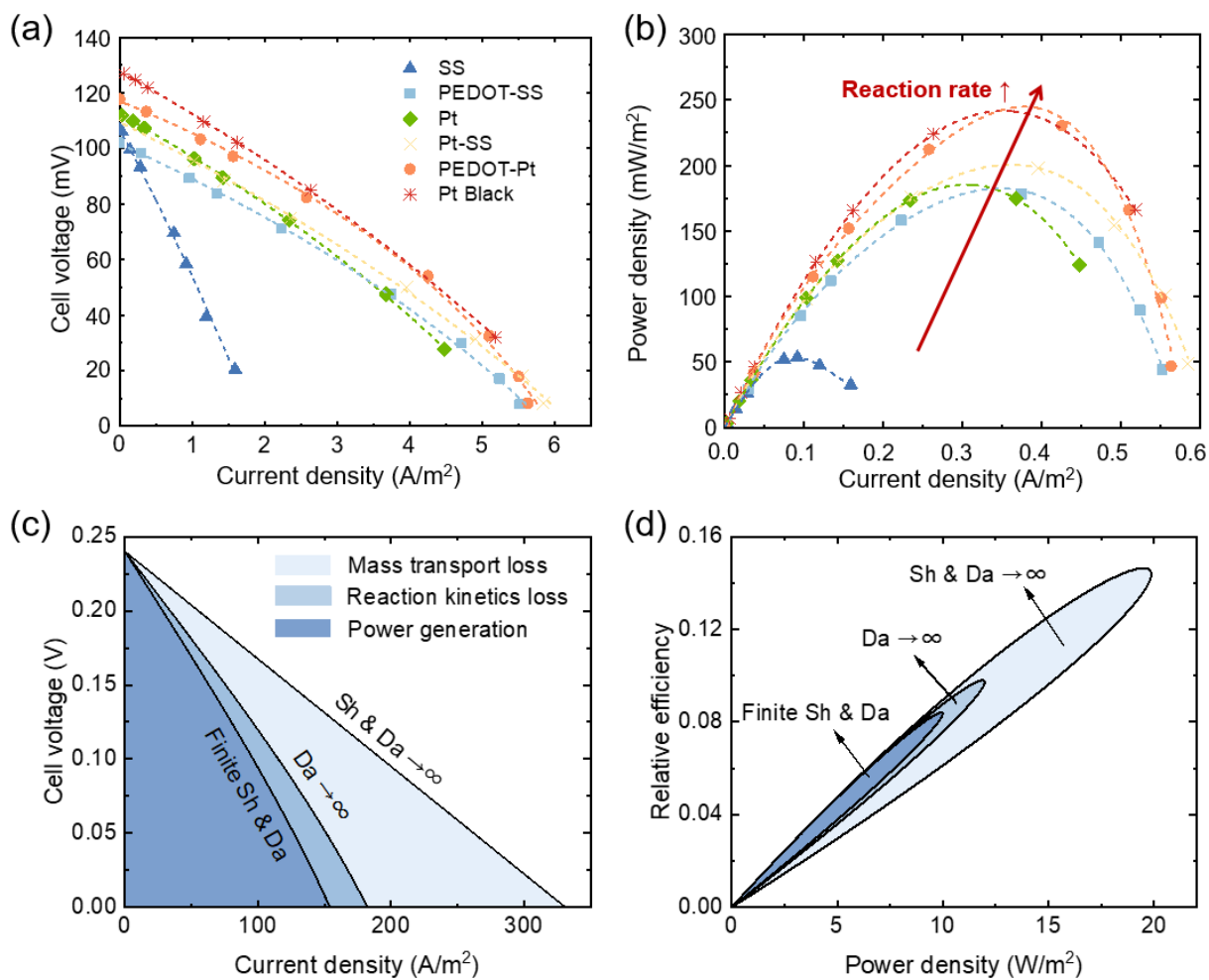

Figure 14. Effect of polarization on the performance of TGCs. (a) Voltage-current curve and (b) power density with different electrodes: Pt Black (asterisks), PEDOT-Pt (circles), Pt (diamonds), Pt-SS (crosses), PEDOT-SS (squares), and SS (triangles). Readapted from ref.[101], Copyright 2014 with permission from RSC. (c) Discharging voltage vs. current density when various losses are considered, and (d) the corresponding loops of efficiency vs. power density at different discharging currents.

This section discusses the thermodynamic analysis of TGCs and provides a modified $Z_{TGC}$ factor that can incorporate the overpotential losses in the electrode-electrolyte interface regions. Typically, the discharging current of TGCs (a few mA/cm$^2$) at optimal efficiency is not large enough to induce strong nonlinear voltage-current relations, which allows us to simplify the computation of overpotential losses by linearizing the Butler-Volmer equation (see Supporting Information S4). As a result, the overall resistance $R_s$ of TGCs can be described by a series of resistances that are associated with different polarizations (Fig. 15a):

$$R_s = R_{ct}^H + R_{mt}^H + R_{ct}^C + R_{mt}^C + R_\Omega, \tag{26}$$

where $R_\Omega$ denotes the total ohmic resistance, $R_{ct}^H$ and $R_{ct}^C$ denote charge transfer resistances at the hot and cold electrodes respectively; $R_{mt}^H$ and $R_{mt}^C$ represent the effective resistances due to mass transfer polarization at the hot and the cold electrodes, respectively. $R_\Omega$ is the total ohmic resistance of the electrolyte and the electrode. These resistances can be individually quantified through electrochemical characterizations. First, the ohmic resistance can be measured by conducting electrochemical impedance spectroscopy (EIS) of the TGC, where the intercept of the impedance curve with the real-axis in the Nyquist plot at high frequencies is the ohmic resistance of device[102]. The charge transfer resistance can also be estimated from the diameter of the semicircle of EIS. Another way to determine the charge transfer resistance is by obtaining the exchange current and the rate constant from the tafel plot and then calculating the charge transfer resistance using Eq. (S31) of Supporting Information. Finally, the mass transfer resistance can be estimated by fitting the low-frequency response of EIS with an equivalent circuit containing the Warburg impedance[103]. If the voltage-current relation of the TGC is linear, the total resistance can be directly estimated using the ratio between open-circuit voltage and short-circuit current.

By linearizing the Butler-Volmer equation, we derived the analytical expression of the maximum efficiency of TGC (see details in Supporting Information S5):

$$\eta_E = \frac{\sqrt{Z_{TGC}\bar{T}+1}-1}{\sqrt{Z_{TGC}\bar{T}+1}+\frac{T_C}{T_H}} \cdot \frac{\eta_C}{1+2/\text{Bi}}, \tag{27}$$

where Bi is the Biot number quantifying the thermal coupling between TGC and the heat reservoirs, $\bar{T} = (T_H + T_C)/2$ is the average temperature. The modified $Z_{TGC}$ is defined using the total resistance and thermal conductance of the TGC devices:

$$Z_{TGC} = S_{tg}^2/(R_s K), \tag{28}$$

which incorporates not only the electrolyte properties but also the properties of the electrodes and the interfaces. Here $K$ is the total thermal conductance of the TGC device. By taking the limit of $\text{Bi} \to \infty$, the efficiency under boundary conditions of fixed temperatures can be obtained as: $\eta_E = \frac{\sqrt{Z_{TGC}\bar{T}+1}-1}{\sqrt{Z_{TGC}\bar{T}+1}+\frac{T_C}{T_H}}\eta_C$, which is similar to the efficiency expression of solid-state thermoelectric generators. Figs. 15b-c show that the analytical expression of efficiency-based modified $Z_{TGC}$ can capture the effect of dimensionless mass transfer rate and the reaction rate. In Fig. 15, the mass transfer rate is nondimensionalized to the Sherwood number $\text{Sh} = ML/D$, where $M$ is the mass transfer coefficient, $L$ is the characteristic length and $D$ is the diffusion coefficient. The electrochemical reaction rate is nondimensionalized as the Damköhler number $\text{Da} = k_0 A_e L^2/D$, where $k_0$ is the rate constant and $A_e$ is the specific area per unit volume. The analytical efficiency based on redefined $Z_{TGC}$ shows excellent agreement with the numerical solution (Supporting Information S6), while the $Z$ factor based on electrolyte properties severely overestimates the relative efficiency. The efficiency map showing full dependence on Da and Sh is also given in Fig. 15d. The modified $Z_{TGC}$ allows for straightforward thermodynamic analysis of TGC devices.

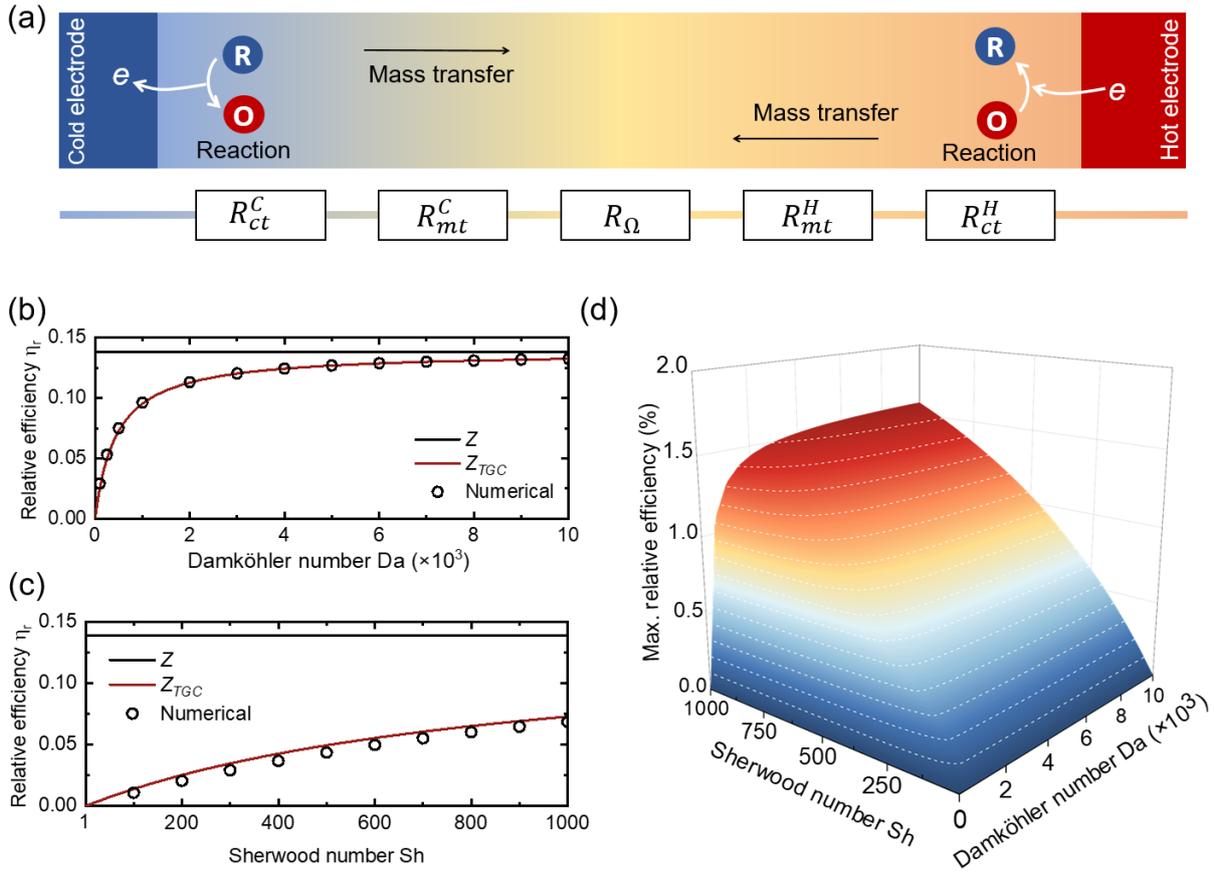

Figure 15. (a) The effective circuit of a TGC device. (b) Maximum relative efficiency $\eta_r = \eta_E/\eta_C$ as a function of Damköhler number Da at the limit of Sh → ∞ (c) The dependence of $\eta_r$ on Sherwood number Sh at the limit of Da → ∞. The original definition of $Z$ is overestimated when Da and Sh are small. (d) Relative efficiency map showing dependence on Da and Sh.

**Electrode engineering and device design of TGCs.** The modified thermoelectric factor $Z_{TGC}$ derived in the previous section indicates that TGCs with smaller equivalent reaction kinetics resistance $R_k$, mass transfer resistance $R_m$, and thermal conductance $K$ have higher energy conversion efficiency. This section discusses methods of engineering the surface activity, morphology, and porous structure of electrodes for reducing $R_k$ and $R_m$, as well as strategies to design TGC devices with reduced thermal conductance $K$ and $R_m$.

*Surface activity.* The surface activity determines the discharging current of TGC devices and can be quantified by the reaction rate constant $k_0$, which is determined by the activation barrier $\Delta E_a$ according to the Arrhenius law $k_0 = A_c e^{-\Delta E_a/\mathcal{R}T}$ [102], with $A_c$ denoting the proportional constant. In TGCs, the activation barrier can be further decomposed to the wetting barrier $\Delta E_{a,w}$ and the electron transfer barrier $\Delta E_{a,e}$. Introducing catalyst is an effective and commonly used way to reduce the electron transfer barrier $\Delta E_{a,e}$ [104-105], such as introducing noble metal coatings [106], embedding noble metal/polyaniline nanoparticles to the electrode surface [107-108], and doping with nitrogen/boron [109] as active sites. On the other hand, reducing the wetting barrier is of equal significance to achieve high discharging current, because unwetted regions of the electrode cannot provide the Faradaic currents necessary for redox reactions. A common practice to increase wettability is by oxidizing the surface of carbon electrodes to introduce hydrophilic functional groups [110]. Zhang *et al.* [110] conducted a comprehensive study on the performance of TGC devices based on various carbon materials, including activated carbon cloth, multiwall carbon nanotubes (MWNTs) foam and sheets, and graphene sheets. Interestingly, the activated carbon cloth, which is the routinely used electrode, has even higher performances than nano carbon materials with high specific surface areas such as MWNTs or graphene sheets. This is owing to the abundant oxygen-functional groups and the exceptional wettability of the activated carbon cloth, while MWNTs and graphene are hydrophobic. Compared with carbon materials, recent developments of MXene-based electrodes have the advantages of intrinsic hydrophilicity as well as the high surface activity of transition metal atoms [111]. Recently, Wei *et al.* [108] developed a ternary composite electrode based on functionalized MXene $Ti_3C_2$ nanosheets, single-wall carbon nanotubes, and polyaniline (PANI), as shown in Fig. 16a. The MXene nanosheets containing hydrophilic terminations serve as active sites for the redox reaction, while the carbon nanotubes provide a porous and electron-conducting framework. PANI particles randomly distributed in the electrode can provide secondary amine functional groups, which are catalysts for redox reactions. As a result, the MXene-based composite material showed a power density of ~11.7 µW/cm$^2$ (Fig. 16b), significantly higher than noble Platinum electrodes.

*Morphology of electrode surfaces.* In addition to surface functionalization, the morphology of the electrode surface is another factor determining the wetting barrier. Increased surface roughness typically results in a small contact angle [112-113] and a higher discharging power density [114-115]. Chemical surface treatments including etching [116-118] and electrodeposition [119-120] can effectively enhance the surface roughness, due to the formation of nanoscale cavities (Fig. 16c). Yu *et al.* [118] showed that the power density can be improved from 0.06 W/m$^2$ to 0.45 W/m$^2$ after electrochemical etching with cyclic voltammetry. Li *et al.* [117] prepared Cu foil electrodes with hierarchical pore structure and showed an order-of-magnitude increase in power density from 29.9 mW/m$^2$ (planar electrode) to 320.7 mW/m$^2$ (Fig. 16d). The hierarchically structured electrodes contain both nanoscale pores for the enhanced surface area as well as micropores and cavities as ion transport pathways to ensure effective supply of reactants. While the nano/microscale structures are usually obtained using chemical processes, larger-scale structures, such as pins and fins, on the electrode surfaces can be prepared through 3D printing [110] or computer-controlled machine tools [118], which can effectively enhance the surface area (Figs. 16e-f). Combining these electrode structuring techniques with chemical treatments could be a future route to prepare electrodes with hierarchically multiscale structures for higher power density.

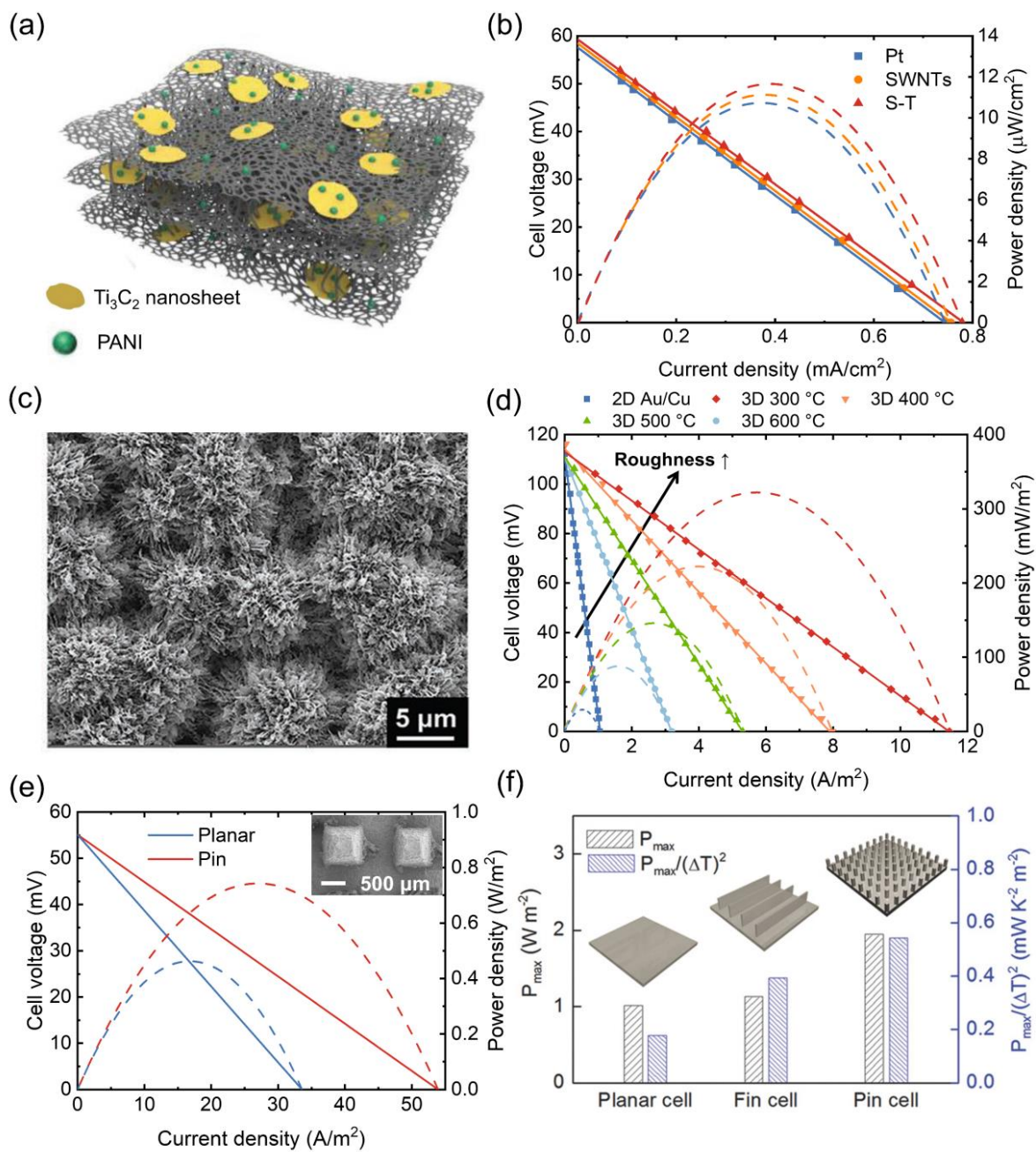

Figure 16. (a) Schematic of a composite electrode containing $Ti_3C_2$ MXene, polyaniline catalysts, and carbon nanotube network. (b) Discharging performance of TGCs based on Pt electrode, single wall carbon nanotube (SWNT) electrodes, and composite electrodes (S-T). Reprinted from ref.[108], Copyright 2022 by Wiley. (c) SEM image of the Cu surface after etching and the corresponding (d) Voltage and the power density of Cu electrodes at different discharging currents at planar (2D) Au on Cu electrode and electrode with 3D morphology prepared at different temperatures (e) Power density improvement by pin structures, with the inset figure showing SEM image., Reprinted from ref. [117], Copyright 2022 by Wiley. (f) Comparison of the performance of planar electrode, fin electrode, and pin electrode. Reprinted from ref.[110] Copyright 2017 by Wiley.

*Device configurations.* Structural configurations of the TGC devices could significantly affect efficiency and power density. For example, the orientation of the TGC device is a factor affecting the discharging performance, due to the natural convection of liquid electrolytes driven by temperature difference across the cell. If the hot electrode is placed over the cold one, natural convection is suppressed resulting in a decreased power density due to the increased mass transfer resistance $R_m$. Kang et al. [121] measured the power output of the three electrode orientations, including hot-over-cold, cold-over-hot, and vertical electrodes (Fig. 17a), and found that both the cold-over-hot configuration and the vertical configuration showed higher power density due to natural convection of electrolytes (Fig. 17b). However, the orientation dependence vanishes if the electrode gap size is smaller than 1.5 cm, because in the small gap the mass transfer is dominated by ion diffusion in the small gap [42].

Although natural convection promotes mass transfer, it also enhances the unwanted heat transfer across the device. However, reducing distances between the electrodes also increases the challenges in maintaining the temperature difference across the cell. Instead of using a single TGC, a possible way is to stack thin TGC devices with the two reservoirs coupled to the two ends. However, such stacked TGC devices would not increase the extracted power despite the increased electrode area, because the temperature difference of the reservoir is fixed, and the internal resistances would increase with extra electrodes. This stack configuration can enhance the efficiency due to the extra thermal resistances (Fig. 17c-d) until the increased activation overpotential and resistive losses become dominant again and decrease the efficiency. Developing thermal separators can be another effective method to improve both the efficiency and the power density, since thermal separators can suppress the unwanted heat transfer and maintain a larger temperature difference across the cell. Ideal separators should be electrically insulative and should have a porous structure for efficient ion transport, and a low backbone thermal conductivity to minimize parasitic heat conduction. Zhang et al. [110] prepared cellulose-based sponge thermal separators and showed an improved temperature difference maintained across the cell, resulting in a two-fold increase in power density.

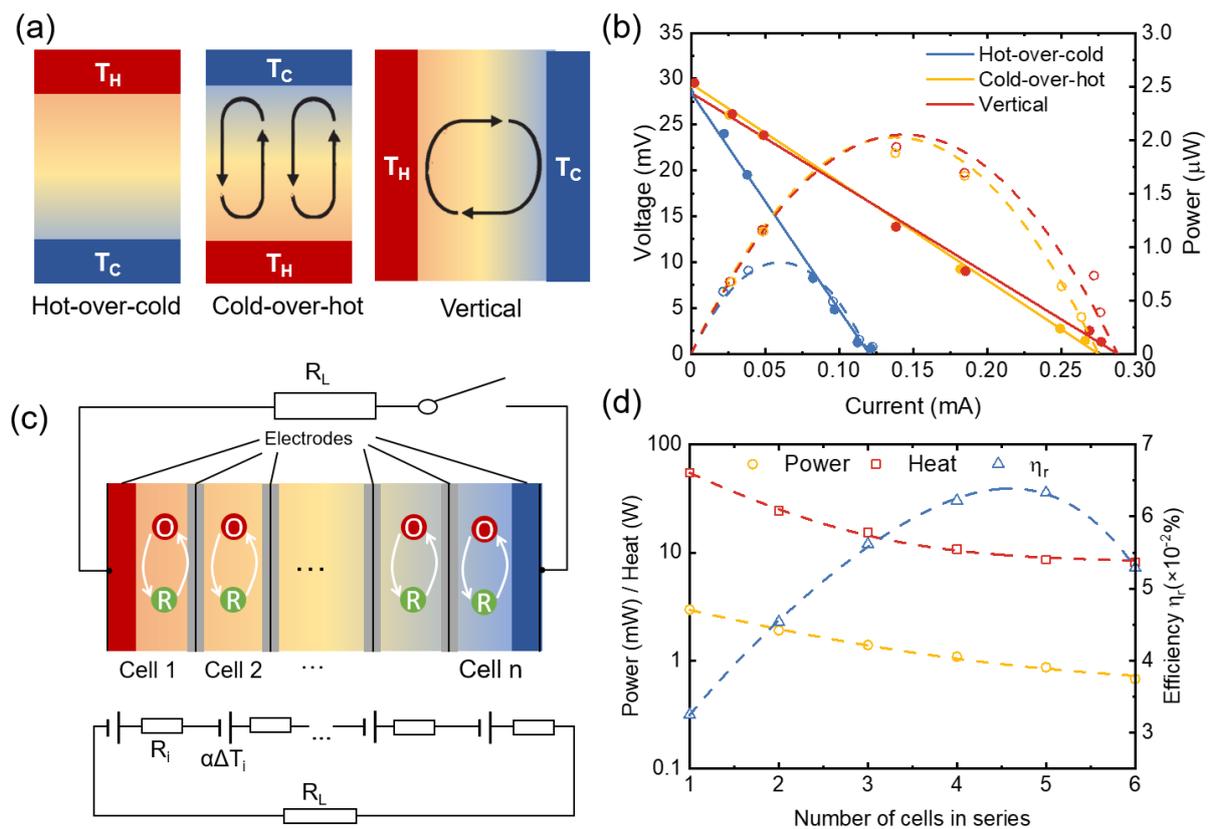

Figure 17. Adjusting electrode orientation and gap size for improving power density and efficiency of TGCs. (a) Natural convection in TGCs with different orientations: hot-over-cold, cold-over-hot, and vertical electrode. (b) The discharging curve and power density with different cell orientations. Readapted from ref. [121], Copyright 2012 by Wiley. (c) Stacked TGCs and the equivalent circuit of discharging. (d) Relative efficiency, power, and input heat in TGCs with stacking electrodes. Readapted from ref. [122]. Copyright 2013, reproduced with permission from Springer Nature.

- **HYBRID I-TE DEVICES**

  Hybrid i-TE devices are integrated devices of different types for enhanced performances or multifunctionality and co-generation. For example, hybrid TIC-TGC and TIC-TEG devices have been proposed to improve the effective thermopower or to simultaneously harvest temperature gradients and fluctuations. Alternatively, coupling i-TE devices with liquid cooling systems or desalination systems can generate electricity while achieving cooling or desalination. This section aims to review recent progress in hybrid i-TE devices. We will summarize the recent developments in hybrid TIC-TGC and TIC-TEG with synergistic thermoelectric conversion effects. Hybrid i-TE devices for co-generation of electricity, cooling power, fresh water, and hydrogen are then discussed.

**Hybrid devices with synergistic effects.** Combining different types of thermoelectric energy conversion devices could be a fruitful direction to achieve improved performances beyond the limit of single-type devices. For example, solid-state TEGs and TGCs suffer from low thermopower due to the limited temperature differences for low-grade heat harvesting and are incapable of harvesting temperature fluctuations. On the other hand, TIC can exhibit high thermopower and can harvest on-and-off temperature changes, but has the disadvantage of transient and small discharging current. This part summarizes recent advancements in combining TIC with TGC or TEG devices to improve thermopower and current outputs, or the capability of simultaneous harvesting temperature gradients and fluctuations.

*Hybrid TIC-TGC devices.* As shown in Fig. 18a, the thermodiffusion of redox-inert ions results in an accumulation of net charges at the hot and cold electrode surfaces, and an internal electric field is induced. For a p-type electrolyte, the electric field is pointing from the cold side to the hot side. On the other hand, the thermogalvanic effect generates a voltage due to the shift of electrochemical potential in response to the temperature. Fig. 18b shows the thermogalvanic effect of the p-type $Fe(CN)_6^{3-}/Fe(CN)_6^{4-}$ pair. The reaction entropy change $\Delta s_{rxn} < 0$ is negative when $Fe(CN)_6^{3-}$ is reduced to $Fe(CN)_6^{4-}$, because the $Fe(CN)_6^{4-}$ ion with higher valence has a more ordered solvation shell, which is a lower entropy state than the $Fe(CN)_6^{3-}$ side. As a result, the oxidization of $Fe(CN)_6^{4-}$ to $Fe(CN)_6^{3-}$ at the hotter electrode is thermodynamically favored to minimize the Gibbs free energy, which donates electrons to the electrode and results in a higher electrochemical potential of the hot electrode. At the cold electrode, the chemical balance shifts towards the favored reduction reaction $Fe(CN)_6^{3-} +e \rightarrow Fe(CN)_6^{4-}$, which extracts electrons from the cold electrode and shifts the electrochemical potential at the cold side downward. Such electrochemical potential difference is in the same direction as the thermodiffusion effect of p-type electrolytes. Therefore, if the electrolyte system contains both the redox pair and the thermodiffusive ions with the same direction of thermal voltage, then they will synergistically contribute to the total thermopower and generate a larger thermal voltage with the same temperature difference than a single-type device. This idea was first proposed and realized by Han and Qian et al. [7]. The authors fabricated a hybrid TIC-TGC device based on ionic gelatin containing both redox active $Fe(CN)_6^{4-}/Fe(CN)_6^{3-}$ and thermodiffusive KCl, and observed a giant thermopower of 17 mV/K, with 17.9% of the total thermopower contributed by the redox reaction, and rest by thermodiffusion of ion species. The giant thermopower enabled the generation of a voltage as high as 2 V from body heat using only 25 i-TE unipolar legs serially connected in an S-shape. However, the ion diffusivity in gelatin is much smaller than that in aqueous solution [123], resulting in limited power density and efficiency for the hybrid TIC-TGC devices. Li et al. [124] proposed an aqueous electrolyte-based asymmetrical TIC-TGC device using $Zn(CF_3SO_3)_2$ electrolyte, with Zn as the anode but $VO_2$ as the cathode (Fig. 18b). Due to the different thermal sensitivities of the asymmetric electrodes, contributions to the thermopower by redox reactions reach ~34%, much higher than the symmetrical TIC-TGC device by Han and Qian *et al* [7]. The thermopower of the asymmetric hybrid TIC-TGC device shows a high thermopower of 12.5 mV/K, and an efficiency of 7.25% relative to the Carnot efficiency, nearly three orders of magnitude higher than the ionic gelatin due to the higher contribution from the redox pair and higher ion diffusivity in aqueous solvents.

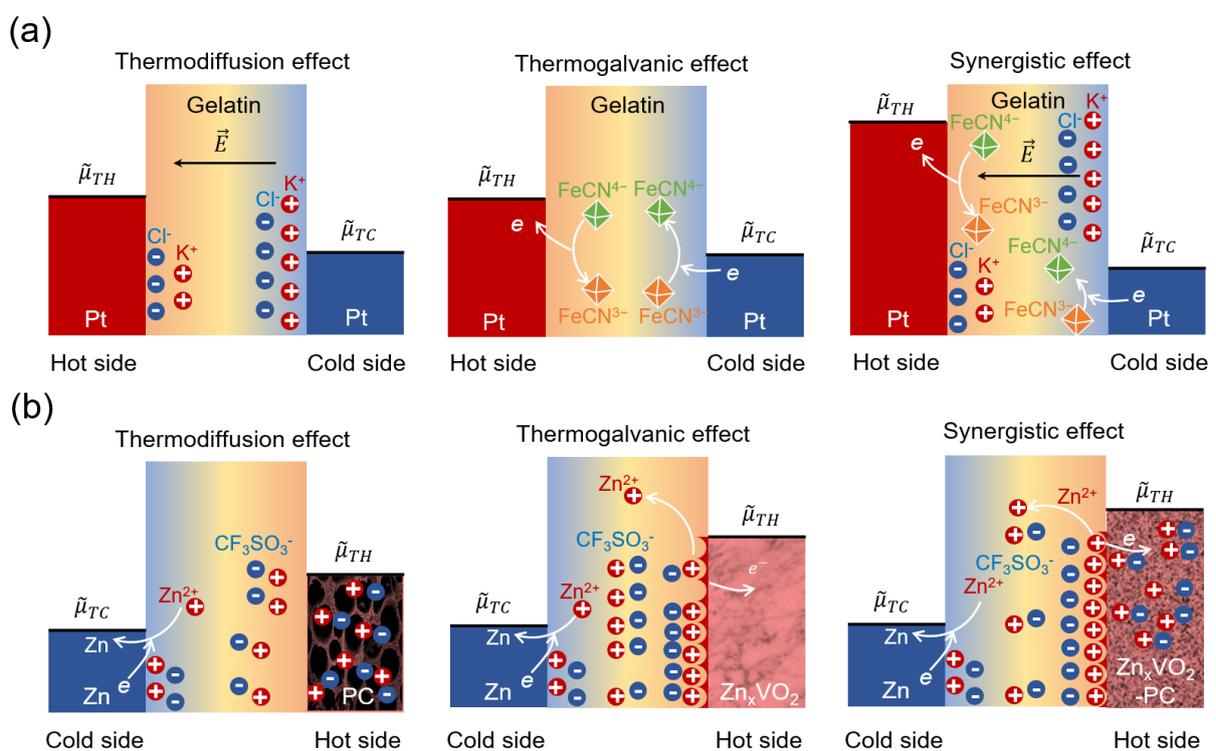

Figure 18. Hybrid TIC/TGC devices. (a) Synergistic contributions to the thermopower by the thermodiffusion of $K^+$ and $Cl^-$ and by thermogalvanic effect of the redox pair $Fe(CN)_6^{3-}/Fe(CN)_6^{4-}$ (denoted as $FeCN^{3-/4-}$) [29]. (b) Asymmetric TIC/TGC device, with $Zn^{2+}/Zn$ as the anode, $Zn^{2+}/Zn_xVO_2$ as the cathode, and $Zn(CF_3SO_3)$ as the ion provider for thermodiffusion effect [124], with PC denoting Porous Carbon.

*Hybrid TIC-TEG devices.* Simultaneously harvesting temperature gradients and temperature fluctuation is of practical importance because low-grade heat sources are usually fluctuating. Solid-state thermoelectric devices can generate continuous and stable voltages while ionogel-based TICs can produce extra power from temperature fluctuations. Recent efforts combining TICs with solid-state TEGs have demonstrated the capability of simultaneous harvesting of temperature gradients and fluctuations from low-grade heat sources. A simple integration of TIC with TEGs is demonstrated by Cheng *et al.* [125], where the PVDF-HPF ionogel loaded with ionic liquid EMIM:DCA and the TE material $Bi_{0.4}Sb_{1.6}Te_3$ are in contact and share the same pair of electrodes. Under temperature gradient, both ions in the gel and holes in the $Bi_{0.4}Sb_{1.6}Te_3$ layer drift from the hot side to the cold side due to the ionic and electronic Seebeck effects. However, due to the much smaller thermopower of $Bi_{0.4}Sb_{1.6}Te_3$ than that of the ionogel, the net charge accumulated at the electrode surfaces is also smaller under the same temperature gradient. When the ionogel and $Bi_{0.4}Sb_{1.6}Te_3$ are in contact (Fig. 19a), electrons or holes in $Bi_{0.4}Sb_{1.6}Te_3$ would migrate towards the ionogel/$Bi_{0.4}Sb_{1.6}Te_3$ interface and the electrode surfaces for charge balance. As a result of this charge redistribution, the output voltage of the hybrid TIC-TEG device is significantly improved compared with the single TEG device. Under a fluctuating temperature, the hybrid TIC-TEG device can generate electricity in five stages (Figs. 19c-e). In the first stage, both the cations and the holes migrate towards the cold side driven by the temperature gradient, leading to a rapid voltage increase along the lateral direction. Second, electrostatic equilibrium across the ionogel/$Bi_{0.4}Sb_{1.6}Te_3$ interface starts to establish. Due to the different Seebeck potentials in the two layers, some cations (in the liquid-state layer) and electrons (in the solid-state layer) migrate towards the contact between the two layers, forming an electric double layer and resulting in potential drops in Stage II, until interface charge equilibrium is established and the thermal voltage stabilized in Stage III. In Stage IV, the heat source is removed, and the accumulated ions or holes relax, resembling the discharging of a capacitor [126]. The sign of voltage is reversed because the interface is already polarized in Stage III which overcompensates the electrolyte charge after the relaxation of ion concentration. Finally in Stage V, all charge carriers are fully relaxed to equilibrium, and the voltage returns to zero. The output power of the TIC-TEG hybrid device is higher than a single TIC and TEG device by nearly 34% [125]. In addition to the simple coupling of TICs with solid-state TEGs, one possible way for further improving the performance is shown in Fig. 19b, where the ionic channel and the electron/hole channel are interpenetrated to maximize the coupling interface. For example, by embedding a carbon nanotube network inside the ionogel [127], the volumetric power density of the TIC/TEG hybrid device can achieve 21 mW/m$^3$ is much larger than that (8 mW/m$^3$) from the ionogel without SWNTs.

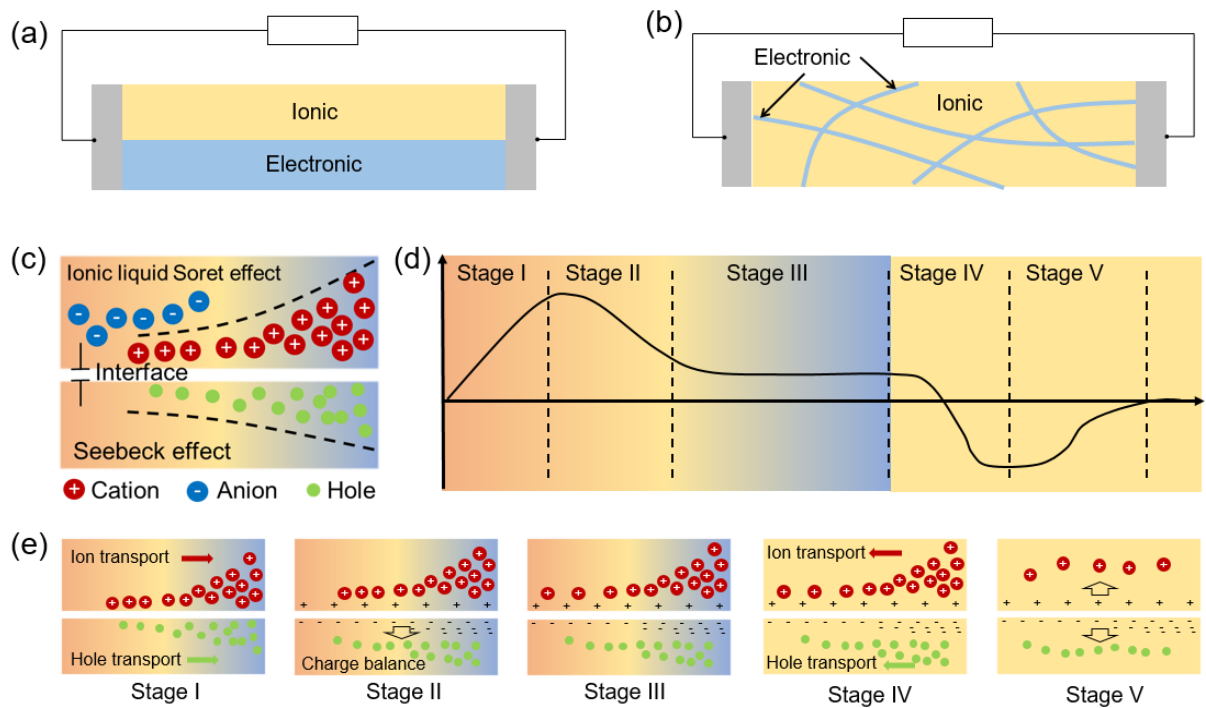

Figure 19. Hybrid TIC/TEG devices developed (a) by simply coupling the i-TE material with electronic TE material through a contact interface and sharing common electrodes, and (b) by interpenetrating the i-TE material with electronic conducting channels. (c) Working principles of the coupled Seebeck and Soret effect across the electrolyte/semiconductor interface under temperature gradient. (d) The thermovoltage profile during the five stages of operation. (e) Schematics of charge migration and balancing in the five stages.

**Hybrid devices for co-generation.** In applications such as building or electronics cooling and solar-driven desalination, a tremendous amount of low-grade heat is generated and dissipated into the environment. Developing devices that can harvest low-grade heat energy and deliver cooling power simultaneously is a promising way to reduce carbon emissions for buildings and data centers. This part discusses recent progress in integrating low-grade heat harvesting i-TE devices with liquid cooling systems or desalination devices for improved efficiency and co-generation of electricity, cooling power, fresh water, and hydrogen.

*Cooling power-electricity co-generation.* Tremendous heat energy wasted in forced liquid cooling systems for electronic devices [128-132] can be recovered. Flow TGCs provide a potential way for recovering low-grade heat energy in forced liquid cooling systems [132-134]. In flow cells, electrolyte coolant flows through a channel composed of two electrodes to dissipate the heat generated, and simultaneously triggers a redox reaction for power generation (Fig. 20a). Kazim *et al.* [132] fabricated a flow TGC for cooling power and electricity co-generation and achieved a power density of 50 mW/m$^2$ with a heat transfer coefficient of 450 W/(m$^2$·K). Ikeda *et al.* [133] performed a comprehensive study on the geometric effects of the flow TGC microchannels for cooling power and electricity co-generation. Interestingly, although embedding fins into the electrode increases the surface area, the output power is decreased due to the deteriorated mass transfer caused by the flow impedance of the fin structures, as shown in Fig. 20b.

*Freshwater-electricity co-generation.* Solar water evaporation has drawn intensive interest in recent years for desalination due to its cost-effectiveness [135-136]. The rapid interfacial evaporation would generate gradients of salinity and temperature, which are not effectively utilized. It is known that reverse electrodialysis (RED) can harvest the salinity gradient [137-138], but the performance of RED is limited by the concentration polarization [139-140]. Fig. 20c shows an example of RED based on the KCl electrolyte with an Ag/AgCl electrode. At the high concentration side, the excess positive charges could accumulate due to the consumption of $Cl^-$ when the Ag electrode is oxidized to AgCl. While there could be cation vacancies for charge balance in TGCs as well, for example, when $I_3^-$ is reduced to $3I^-$ ions, two $K^+$ ions (cation vacancies) need to be supplied for charge neutrality. In viewed by the same cation surplus and vacancies, combining RED and TGC can reduce the concentration polarization while simultaneously harvesting the temperature gradient and salinity gradient. As shown in Fig. 20d, Wang et al. [141] realized the hybrid RED-TGC system, where a TGC is sandwiched by two RED half cells, separated by two composite electrodes. $K^+$ cation surplus generated by the high concentration half-cell is supplied to the hot side of TGC to reduce the cation shortage and therefore reduce the electrode polarization for both devices. By combining the RED-TGC system with the solar evaporator (Fig. 20e), excellent power-water co-generation performance is achieved with a power density of 1.1 $W/m^2$ with a water production rate of 1.4 kg $m^{-2}$ $h^{-1}$ (Fig. 20f) under one sun. Power contributions from the RED and TGC in the hybrid device are 0.85 $W/m^2$ and 0.26 $W/m^2$ respectively, which are significantly higher than a single RED (0.5 $W/m^2$) or a single TGC (2.2 $mW/m^2$).

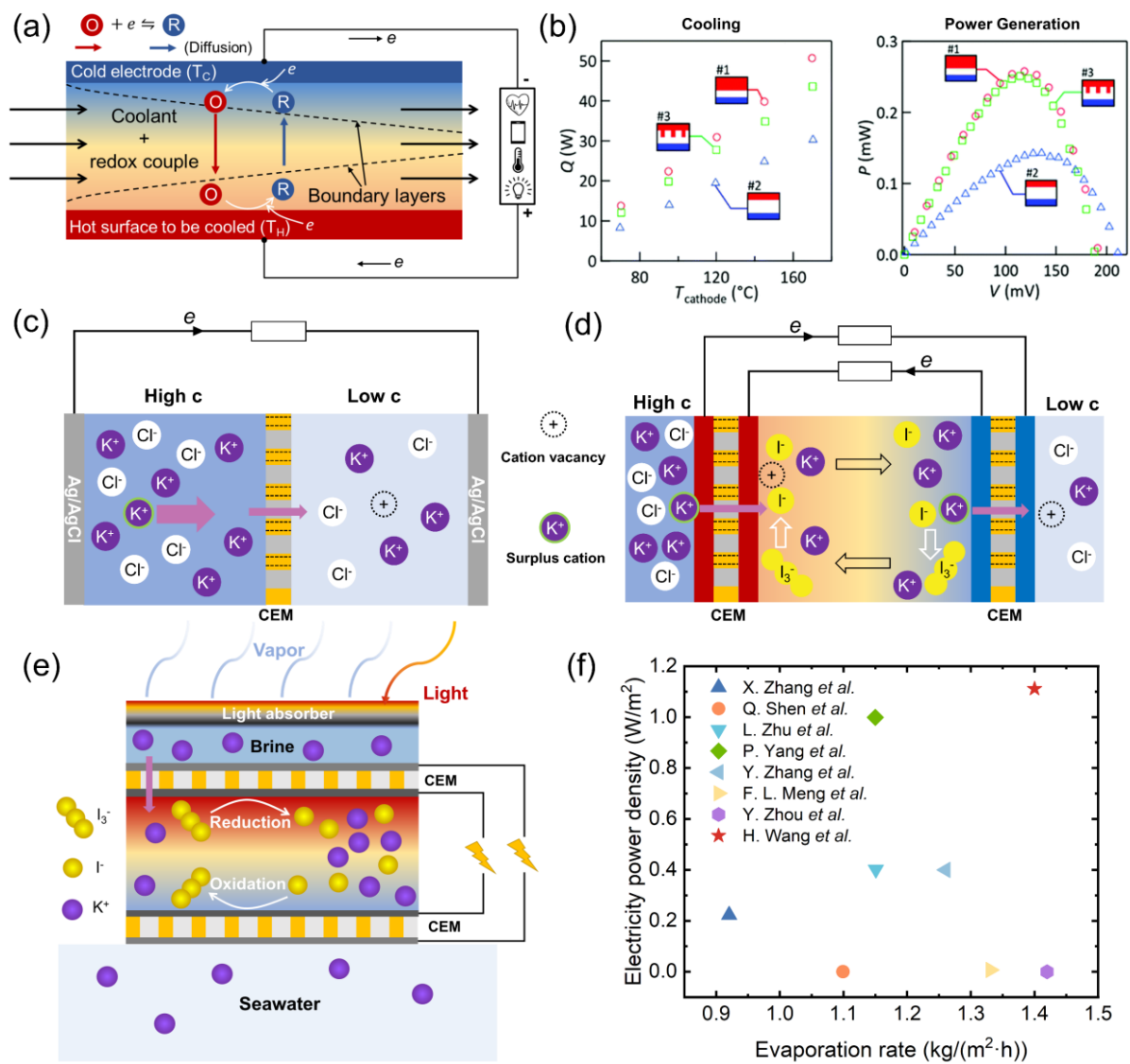

Figure 20. Hybrid devices for cooling power-electricity and freshwater-electricity co-generation. (a) Schematic of the cooling channel and the flow TGC hybrid device. (b) Cooling and power generation of the hybrid device. Reprinted from ref. [133], Copyright 2019 with permission from RSC. Schematic of the ion transport in (c) RED and (d) the TGC-RED hybrid device. (e) The hybrid device combining solar-driven interfacial evaporation with TGC-RED [141] for the co-generation of water and electricity. (f) Evaporation rate and electricity power density values of the hybrid systems with TGC-RED, compared with results by Zhang et al. [142], Shen et al. [143], Zhu et al. [144], Yang et al. [138], Zhang et al. [145], Meng et al. [146], and Zhou et al. [147]. Readapted from ref. [141], Copyright 2021 by Wiley.

*Hydrogen-electricity co-generation.* Recently, Wang et al. [148] reported a strategy to enhance the power density and efficiency of TGCs while simultaneously achieving hydrogen production, by embedding oxygen/hydrogen evolution photocatalysts (OEP/HEP) to the hot and code sides of TGCs (Fig. 21a), respectively. The working principles of the photocatalytic enhancement effect on TGC are shown in Fig. 21b-c. Under light illumination, electrons and holes are generated in the oxygen vacancies in $WO_3$ particles ($O_v$-$WO_3$) as OEP (Fig. 21b). The conduction band minimum (CBM) of the $O_v$-$WO_3$ is above the redox potential $Fe(CN)_6^{3-/4-}$, such that the photo-generated electrons favor the reduction of $Fe(CN)_6^{3-}$ into $Fe(CN)_6^{4-}$. The increased concentration of $Fe(CN)_6^{4-}$ on the hot side is beneficial for improving $S_{tg}$. On the other hand, the valence band maximum (VBM) of $O_v$-$WO_3$ is below the water-oxidizing ($H_2O/O_2$) potential, such that the holes can be accepted by $H_2O$ molecules and $O_2$ is generated. Similarly, the CBM of sulfur vacancies in $ZnIn_2S_4$ ($S_v$-ZIS) as HEP is above the water-reducing ($H^+/H_2$) potential, and the VBM is below the redox potential of $Fe(CN)_6^{3-/4-}$ (Fig. 21c). As a result, the photogenerated electrons result in reduction of water and hydrogen generation, and the holes are accepted by $Fe(CN)_6^{4-}$, resulting in an increased concentration of $Fe(CN)_6^{3-}$ at the cold electrode. The localized high concentrations of $Fe(CN)_6^{4-}$ at the hot anode and $Fe(CN)_6^{3-}$ at the cold cathode resulted in a thermopower as high as 8.2 mV/K, which significantly increased the voltage and power density of TGC under light illumination (Fig. 21d-e) compared with pristine TGCs. A record-high $P_{max}/\Delta T^2$ of 8.5 mW/($m^2 \cdot K^2$) is achieved, and the hybrid TGC device also achieved a solar-to-hydrogen efficiency of up to 0.4%.

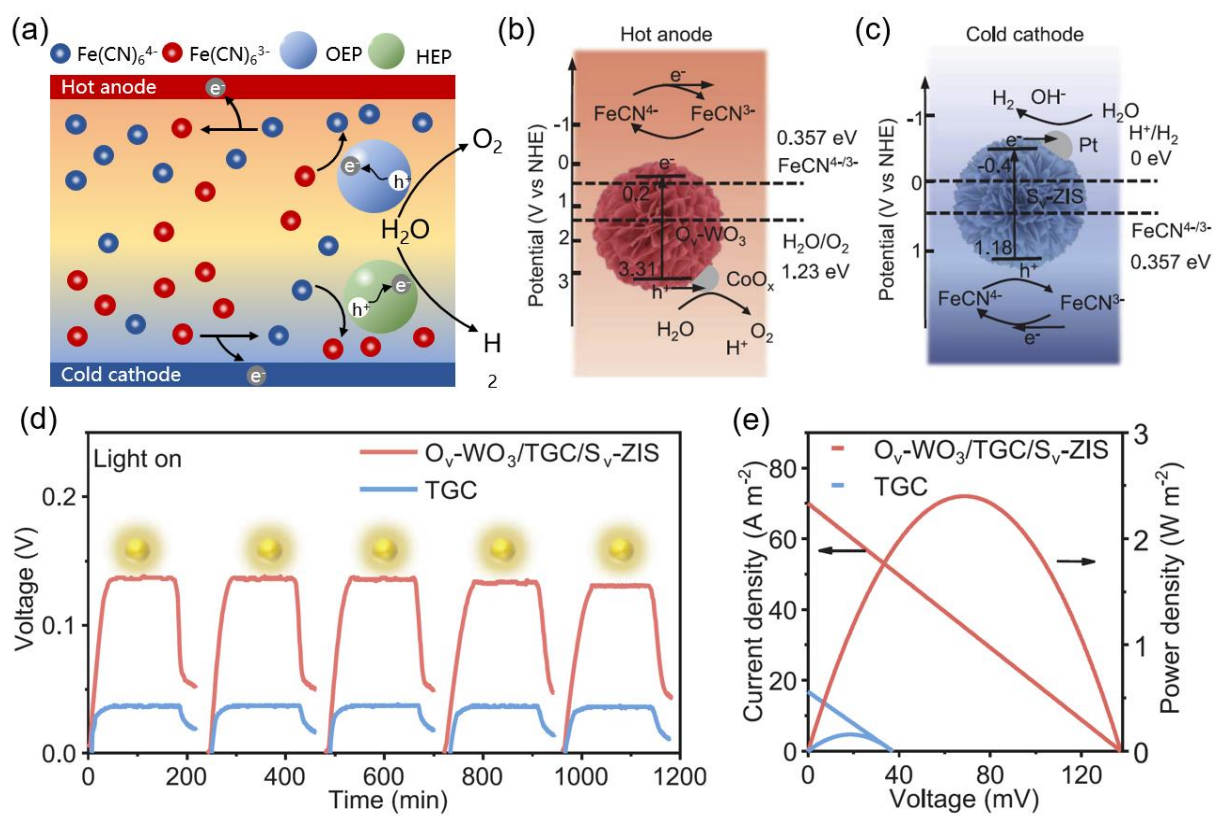

Figure 21. Photocatalytically enhanced TGC for co-generation of electricity and hydrogen. (a) Schematic of the photocatalytically enhanced TGC by embedding oxygen evolution photocatalysts (OEP) and hydrogen evolution photocatalysts (HEP) into the TGC. (b-c) Electrode potential and energy gap for the photocatalysts. Under light illumination, (b) electrons and holes are generated by the OEP ($O_v$-$WO_3$) particles, with the electrons above the $Fe(CN)_6^{3-/4-}$ redox potential accepted by $Fe(CN)_6^{3-}$, and holes below the water oxidizing potential being accepted by water molecules generating $H^+$ and $O_2$; (c) electrons and holes are generated in the HEP ($S_v$-ZIS particles), with electrons above the water reducing potential favoring $H_2$ production and holes accepted by $Fe(CN)_6^{4-}$. (d) Comparison of voltage between the TGC and the photocatalytically enhanced cell $O_v$-$WO_3$/TGC/$S_v$-ZIS. (e) Discharging current, voltage, and power density between the TGC and photocatalytically enhanced hybrid device. From ref. [148]. Copyright 2023, Reprinted with permission from AAAS.

**Performance comparisons.** To conclude the discussions, we summarize the performances of hybrid i-TE devices and make a parallel comparison with TICs, TGCs, and other electrochemical-based techniques for harvesting low-grade heat. Fig. 22 shows a comparison in efficiency and thermopower among different types of i-TE devices. Since thermally regenerative electrochemical cycle (TREC) batteries [8, 10, 95, 149-151] are also based on the temperature-sensitive electrochemical potential, they are also included in Fig. 22. While TICs are featured by the giant thermopower (Fig. 22a), the efficiency remains low compared with TGCs or TREC devices with redox ions. The major limiting reasons are the long thermal charging time, large internal resistances, and the fast decaying output currents. Different from TGCs and TICs that generate a voltage from the spatial temperature difference, TRECs harvest temperature differences in the time domain by charging and discharging at different temperatures. The relatively high efficiency of TREC devices is due to the absence of irreversible heat conduction across the device. Recent developments of Zn-ion-based hybrid i-TE cells[152-153] with synergistic thermodiffusion and thermogalvanic effect demonstrate a feasible route to develop high-performance i-TEs, showing a relatively high thermopower, reasonable efficiencies, and power densities. The TGC with thermosensitive crystallization processes also show much higher power density and efficiency than TGCs based on liquid electrolytes, due to the increased concentration gradient of the reactants across the device[42].

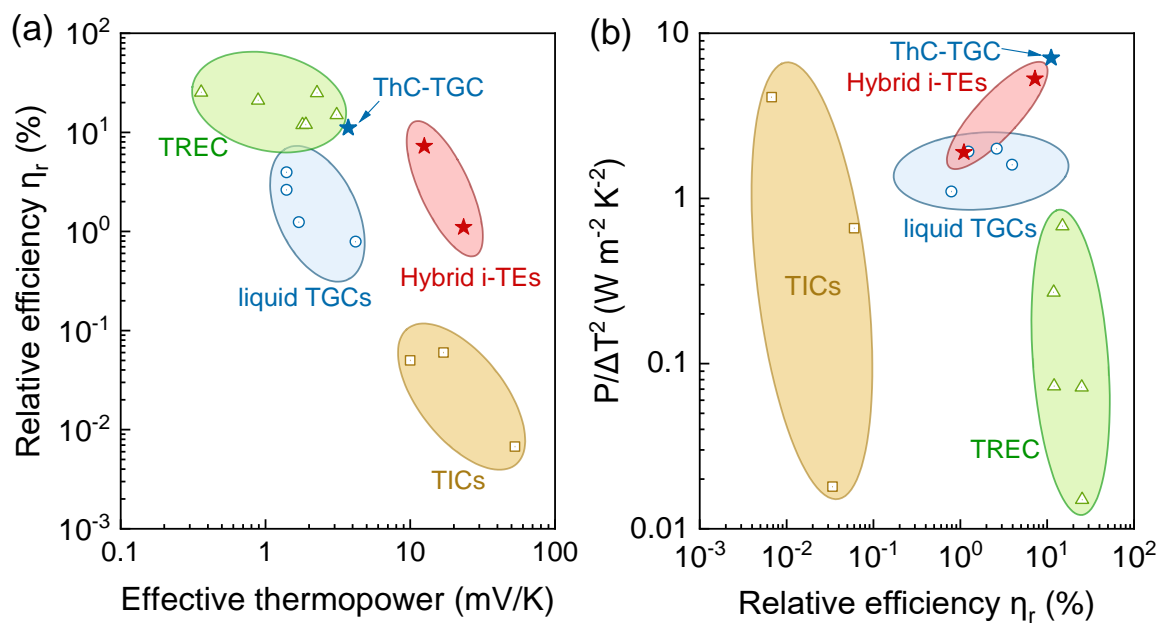

Fig. 22. Performance comparisons of i-TE and TRECs. (a) Relative efficiency *vs* effective thermopower, and (b) Normalized power density $P/\Delta T^2$ *vs* relative efficiency of liquid TGCs [154-157], TGCs with thermosensitive crystallization (ThC-TGC) [42], TICs[23, 30, 158], and hybrid i-TE devices [152-153].

In Fig. 23, we compare the absolute efficiency and power density of TGCs and hybrid i-TE devices with other electrochemical techniques for low-grade harvesting, including vacuum distillation-concentration redox flow battery (VD-CRFB)[159-160], vacuum distillation/membrane distillation-reverse electrodialysis (VD/MD-RED) [161-162], thermolysis-reverse electrodialysis (TL-RED)[163-164], vacuum distillation/membrane distillation-pressure retarded osmosis (VD/MD-PRO)[165-166], thermally regenerative ammonia battery (TRAB)[167-169] and thermally regenerative copper acetonitrile (CuACN) batteries[170]. Comprehensive reviews of these electrochemical methods for harvesting low-grade heat can be found in published literature [15, 171]. TIC devices are not included in the comparison due to their fast decay in output power. The typical power density of liquid TGCs ranges from 1 W/m$^2$ to 10 W/m$^2$, while the efficiency is limited by both the heat conduction across the device and the achievable $\Delta T$ across the device. The record-high efficiency and power densities among i-TE devices were measured in TGCs with thermosensitive crystallization[42]. The recent development of Zin-ion-based hybrid i-TE devices also showed comparable power density and efficiency with TRABs, but still lower than CuACN batteries and VD-CRFBs. However, focusing on improving the thermopower might sometimes deteriorate the performance of i-TE devices. For example, among the two Zinc-ion-based hybrid i-TE devices reported by Zhang's group[152-153], the device with higher power density and efficiency shows a much lower thermopower. Instead of simply increasing the thermopower, future research working on i-TEs should focus on looking for strategies for simultaneous improvement in thermopower, reaction kinetics, and conductivity of the electrolytes.

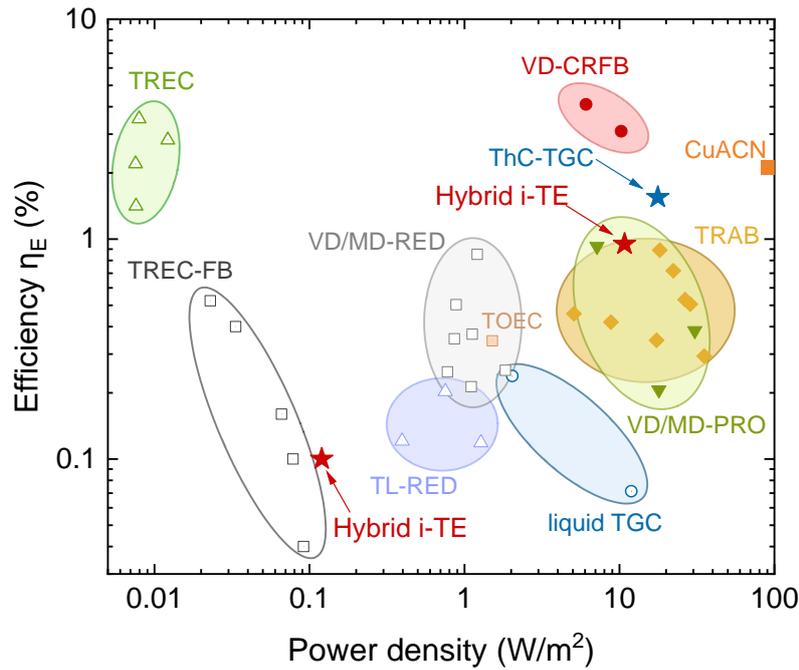

Figure 23. Performance comparisons of i-TE devices with other electrochemical devices with data of different electrochemical techniques taken from refs. [10, 15, 171]. Data for hybrid i-TEs and TGCs with thermosensitive crystallization (ThC-TGCs) were collected from refs. [42, 152-153]. TIC devices are not included due to their low efficiency and fast decay in output power.

▪ **SUMMARY AND OUTLOOK**

In summary, this review article comprehensively discussed the working principles, electrochemistry, and thermodynamics of i-TE devices including TICs, TGCs, and hybrid devices. The thermopower of both TIC and TGCs can be understood from the entropic point of view, while the Soret thermopower $S_{td}$ is related to the Eastman entropy of transfer but the thermogalvanic thermopower $S_{tg}$ is related to the entropy change of the redox reaction. Strategies for improving the thermopower of TICs and TGCs are summarized under the framework of tuning Eastman entropy of transfer or reaction entropy changes, respectively. The thermodynamics of TICs and TGCs are analyzed in detail, and this review redefined $Z$-factors for TICs and TGCs at the device level and derived analytical expressions for efficiencies.

When evaluating the energy conversion efficiency of i-TE devices, the finite thermal charging time of TICs, the overpotential losses especially at the electrode-electrolyte interfaces of TGCs, and the finite heat transfer coefficient must be considered, simplistically reporting the *ZT* factors based on only material properties is meaningless. Recent advancements in electrode engineering and device design for both TICs and TGCs are then summarized from the viewpoint of thermodynamics and ion transport. In addition to simple i-TE devices, hybrid TIC-TGC and TIC-TEG devices with synergistic thermoelectric effects, and hybrid devices for co-generation of electricity, cooling power, fresh water, and hydrogen are reviewed. Finally, we comment on several remaining challenges and possible routes for future development of i-TE devices:

(i) Methods of predicting the Soret and thermogalvanic thermopower are in urgent demand for the development of high-thermopower i-TE materials but remain highly challenging due to the complicated and long-range feature of ion-ion and ion-solvent interactions. Although recent MD simulations have been applied to predict $S_{tg}$ of $Fe^{3+}/Fe^{2+}$ and $Fe(CN)_6^{3-}/Fe(CN)_6^{4-}$ redox pairs [98], it is only applicable to a dilute solution of simple redox pairs and cannot predict thermopower of redox reactions involving chemical species from the electrodes. A computational framework for studying $S_{td}$ in complex electrolytes (*e.g.*, polyelectrolytes, ionic liquids, and gels) is still lacking.

(ii) Simultaneously enhancing the thermopower and ionic conductivity of i-TE materials could be challenging, similar to the case in solid-state thermoelectric materials. The coupling between ionic thermopower and ionic conductivity/diffusivity needs further study to improve the efficiency of harvesting low-grade heat. For example, polyelectrolytes or ionogels show giant $S_{td}$ typically sacrifices ion transport properties due to the increased hopping barrier. For TGCs, the $S_{tg}$ is possibly coupled with the ion transport coefficients, because regulating the solvation structure of redox species might also affect the effective ionic radii or introduce ionic couplings, which could affect the diffusivities.

(iii) Current research mostly focuses on improving the thermopower of either n-type or p-type materials. Nonetheless, thermopower matching between n-type and p-type i-TE would determine the overall performance of an i-TE pair. For example, in contrast to the significant advances in p-type TGCs, such as the $Fe(CN)_6^{4-}/Fe(CN)_6^{3-}$ system, their n-type counterparts with higher effective thermopower needs to be developed.

(iv) According to the thermodynamic analysis in this review, heat transfer between the i-TE device and the heat reservoir should be paid more attention to for higher efficiency of the system. When the i-TE device is coupled with low-grade heat sources, the cold side temperature is usually higher than the air temperature due to the ineffective natural convective heat transfer (low Bi number). Improving the coupling between the i-TE device with the heat reservoirs is crucial when harvesting small temperature differences and fluctuations. A large heat transfer area might be necessary to achieve enough $\Delta T$, therefore, it is also suggested that the heat-transfer-area normalized power densities should also be compared with other techniques when reporting future results[15].

**Quotes for the Review:**

While current research reported great progress in improving ionic thermopower, fundamentals of thermodynamics, electrochemistry, and ion transport in i-TE devices are yet comprehensively investigated despite the pivotal importance of affecting power density and efficiencies.

The thermopower of both TIC and TGCs can be understood from the entropic point of view, while the Soret thermopower $S_{td}$ is related to the Eastman entropy of transfer but the thermogalvanic thermopower $S_{tg}$ is related to the entropy change of the redox reaction.

When evaluating the energy conversion efficiency of i-TE devices, the finite thermal charging time of TICs, the overpotential losses especially at the electrode-electrolyte interfaces of TGCs, and the finite heat transfer coefficient must be considered, simplistically reporting the *ZT* factors based on only material properties is meaningless.

- **AUTHOR BIOGRAPHIES**

**Xin Qian** is currently a tenure-track professor of engineering thermophysics at Huazhong University of Science and Technology (HUST). His expertise includes the transport physics of ions and phonons in energy materials and ultrafast thermal characterizations. Link to personal webpage: https://sites.google.com/view/xin-qian-/

**Zhihao Ma** obtained his Master's degree in engineering thermophysics at HUST and is currently a Ph.D. student at the University of Utah. His research focuses on heat and mass transfer modeling of electrochemical devices and radiative cooling materials.

**Qiangqiang Huang** is currently a Ph.D. student co-advised by Xin Qian and Ronggui Yang in energy and power engineering at HUST. His research focuses on electrochemical materials and devices for harvesting low-grade heat.

**Haoran Jiang** is a professor of mechanical engineering at Tianjin University. His research focuses on heat and mass transfer in electrochemical materials and devices, including redox flow batteries for energy storage, thermal management technologies of batteries, and innovative electrochemical devices for low-grade heat harvesting.

**Ronggui Yang** is a professor of Energy and Power Engineering at HUST and was a professor of Mechanical Engineering at CU Boulder (2006–2019). His expertise includes the fundamentals of heat conduction, radiation, phase-change heat transfer, and the applications of micro/nanotechnologies in energy systems. Link to the webpage: http://x-thermal.energy.hust.edu.cn/english/people/pi.htm

- **ASSOCIATED CONTENTS**

**Supporting Information:** Derivation of time scale of thermal responses, Figure of Merits and efficiencies, stacking electrode models of TICs and TGCs, overpotential resistances of TGCs, and numerical results of figure of merits.

- **ACKNOWLEDGMENTS**

X.Q. acknowledge support from the National Natural Science Foundation of China (NSFC Grant No. 52276065). R.Y. acknowledges financial support from the National Key Research and Development Program of China (Grant No. 2022YFB3803900). H.J. acknowledges support from the National Natural Science Foundation of China (NSFC Grant No. 52106265). The authors declare no conflict of interest.